\documentclass[aps,prl,reprint,superscriptaddress,nofootinbib]{revtex4-2}

\usepackage{amsmath,amssymb,bm}
\usepackage{booktabs}
\usepackage{graphicx}
\graphicspath{{anyon_dispersion_figures_v29/}}
\usepackage[colorlinks=true,allcolors=blue]{hyperref}

\newcommand{\ii}{\mathrm{i}}
\newcommand{\dd}{\mathrm{d}}
\newcommand{\eps}{\epsilon}
\newcommand{\Th}{\Theta}
\newcommand{\Imop}{\operatorname{Im}}

\newcommand{\bk}{{\bm k}}
\newcommand{\bp}{{\bm p}}

\newcommand{\bq}{{\bm q}}
\newcommand{\bP}{{\bm P}}
\newcommand{\bK}{{\bm K}}

\newcommand{\br}{{\bm r}}
\newcommand{\calM}{{\mathcal M}}

\begin{document}

\title{Measuring anyon dispersion with tunneling probes}

\author{Taige Wang}
\affiliation{Materials Research Laboratory, Massachusetts Institute of Technology, Cambridge, MA 02139, USA \looseness=-2}
\affiliation{Department of Physics, Harvard University, Cambridge, MA 02138, USA \looseness=-2}

\author{T. Senthil}
\affiliation{Department of Physics, Massachusetts Institute of Technology, Cambridge, MA 02139, USA \looseness=-2}

\date{\today}

\begin{abstract}
Anyons are usually characterized by their topological data and their fractional quantum numbers under global symmetries. In lattice systems such as fractional Chern insulators (FCI), they are also mobile quasiparticles. Their motion controls the possible ground states of the dilute anyon gas obtained by doping an FCI, including possible superconducting states. We show how tunneling probes can measure this motion. In scanning tunneling spectroscopy, weak disorder produces spatially oscillating quasiparticle-interference patterns whose branches reveal the dispersion of fractionalized constituents. In quantum twisting microscopy, planar momentum-conserving tunneling selects the total momentum of the injected electron, so the continuum thresholds of fractionalized electron spectra encode the dispersion of the constituent anyons. The resulting spectra distinguish compact electron-like excitations, bound anyon molecules, and unbound anyon continuum.
\end{abstract}

\maketitle

\emph{Introduction.---}
Fractional quantum Hall states introduced quasiparticles with fractional charge and fractional statistics~\cite{Laughlin1983,Arovas1984}.  These topological properties are the usual experimental targets: shot noise measures charge, interferometry probes braiding, and thermodynamics probes gaps.  In a continuum Landau level, however, charged anyon motion is naturally quenched by the magnetic-translation algebra.

Fractional quantum anomalous Hall states in twisted MoTe$_2$~\cite{ZengMoTe2,ParkMoTe2} and rhombohedral multilayer graphene~\cite{LuMLG,LuExtendedQAH2025,AronsonFCI2025,WatersPentalayer}, together with lattice fractional Chern insulators (FCI) in graphene--hBN moir\'e~\cite{Spanton2018FCI,TsuiHofstadterUnpublished} and magic-angle graphene~\cite{XieMATBG} under strong magnetic fields, change this situation.  Their anyons can carry crystal momentum in a reduced magnetic Brillouin zone.  Recent theory further suggests that periodic potentials and magnetic-field inhomogeneity can generate dispersive anyon bands in FCIs and related FQAH systems~\cite{Shi2024_doping, SchleithSoejimaKhalaf,IyerAnyonBloch,YanLiSoejimaKhalaf,LeeWangZaletelVishwanathHe2018,WangSongZaletelSenthil2025,LuWuChenMeng2025}.  Beyond charge and statistics, one would therefore like to measure how an anyon moves.

This motion is not a minor detail.  In a doped FCI, anyon masses and bandwidths set the density of states of a dilute anyon gas, while residual interactions determine whether carriers remain isolated, bind into anyon molecules, or condense.  Superconductivity and reentrant anomalous Hall behavior observed nearby in moir\'e systems~\cite{HanRhomboSC,XuMoTe2SC} make these kinetic data a microscopic input to classic anyon-driven superconductivity ideas~\cite{LaughlinAnyonSC,LaughlinScience1988,Chen1989_anyonSC,LeeFisher1989,Halperin1992} and to recent theories of doped fractional anomalous Hall insulators~\cite{Shi2024_doping,Kim2024_anyonSC,shi2025doping,shi2025anyon,NosovHanKhalaf,Pichler2025,WangZaletelSC,FanVishwanathWang2026,shi2025non}.

Scanning probes are natural for this task because weak inhomogeneity converts motion into spatial oscillations.  In ordinary electron systems, Fourier-transform scanning tunneling spectroscopy (STS) uses quasiparticle interference (QPI) to infer scattering wave vectors from energy-resolved LDOS modulations~\cite{Crommie1993,Hasegawa1993,Hoffman2002,McElroy2003,WangLee2003,Capriotti2003,PeregBarneaFranz2003,Hanaguri2010,Roushan2009,Mallet2012,SimonReview,BenaReview,AvrahamReview}.  We propose the analogous STS-QPI measurement for fractionalized electron-addition states.  The clean activation threshold already distinguishes compact electron-like excitations, bound anyon molecules, and unbound anyon continuum through its power law~\cite{Morampudi2017Threshold}.  With weak disorder, the projected impurity potential can scatter one charged constituent while the others remain spectators, so the QPI branches reveal the constituent dispersion.  A complementary route is quantum twisting microscopy (QTM), where a planar graphene probe injects electrons with twist-selected momentum~\cite{InbarQTM,XiaoMATBGQTM}.  Since the FCI electron spectrum contains multi-anyon continuum, the fixed-momentum thresholds measured by QTM provide an independent extraction of the same anyon kinetics.

\begin{figure}[t]
\centering
\includegraphics[width=\linewidth]{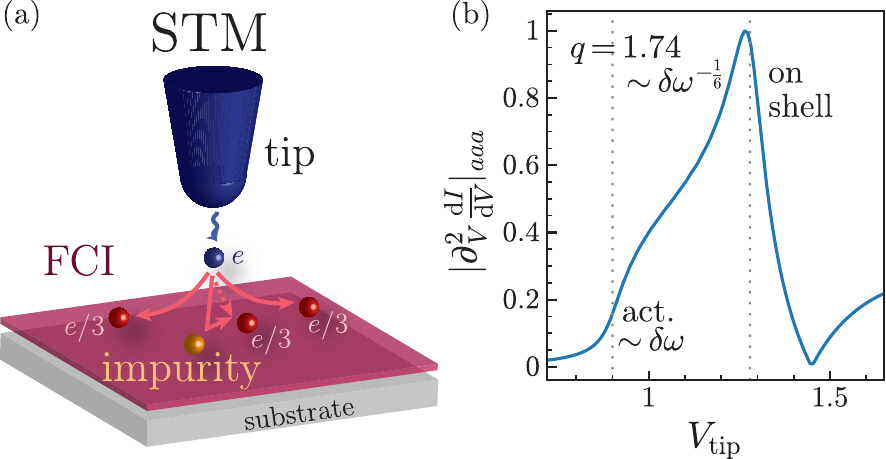}
\caption{STS-QPI geometry and a fixed-$q$ line cut.  (a) An electron injected from the STM tip can enter the FCI as an $aaa$ continuum of three charge-$e/3$ anyons.  A weak impurity transfers momentum $\bq$ to one constituent, and the interference between amplitudes with and without this scattering produces the $\bq$ component of the conductance map.  (b) Numerical fixed-$q$ line cut for the $aaa$ sector.  The first dashed line marks the clean activation threshold $3\Delta_a$, while the second marks the on-shell QPI branch $\omega_{aaa}(q)$.  The indicated powers are those of $\partial_V^2(\dd I/\dd V)$ for Laughlin $\nu=1/3$.}
\label{fig:stm}

\end{figure}

\emph{STM-QPI.---}
We now focus on a Laughlin-like $\nu=1/3$ state.  The fundamental anyon is denoted $a$ and carries charge $e/3$.  A possible charge-$2e/3$ molecule is denoted $b$, and a compact charge-$e$ excitation is denoted $e$.  The physical electron-addition spectrum can therefore contain the sectors
\begin{equation}
  e,\qquad ba,\qquad aaa.
\end{equation}
Which sector is lowest is a microscopic energetic question, controlled for example by gate screening and by the anyon-molecule hierarchy~\cite{WangZaletelMolecules,XuClusters,GattuJainMolecular,LiNosovKhalaf}.  When explicit formulas are needed, any mobile object or molecule $s=a,b,e,\ldots$ is assigned the low-energy dispersion
\begin{equation}
  E_s(k)=\Delta_s+{k^2\over 2M_s}.
  \label{eq:generic-dispersion}
\end{equation}

In the Tersoff-Hamann limit, positive-bias STS measures
\begin{equation}
  g(\br,\omega)\equiv {\dd I\over \dd V}(\br,\omega)
  \propto A_e(\br,\omega)
  =-{1\over\pi}\Imop G_e^R(\br,\br;\omega).
\end{equation}
Here the local electron creates the state $\psi_e^\dagger(\br)|0\rangle$, so STM always measures the physical electron propagator
\begin{equation}
  G_e^R(\br,\br';\omega)=
  \langle 0|\psi_e(\br)
  {1\over \omega-(H-E_0)+\ii0^+}
  \psi_e^\dagger(\br')|0\rangle .
  \label{eq:electron-resolvent}
\end{equation}
This propagator can be decomposed into different charge-$e$ sectors.  We use $\lambda=e,ba,aaa,\ldots$ to label such a sector, while $s=a,b,e,\ldots$ labels a constituent species or mobile object.  Let $|n,\bK;\lambda\rangle$ be an exact eigenstate in sector $\lambda$.  The sector-resolved contribution is
\begin{align}
  G^R_{e,\lambda}(\omega)
  &=\sum_{\bK,n}
  { |\calM^\lambda_n(\bK)|^2 \over
  \omega-E^\lambda_n(\bK)+\ii0^+},
  \label{eq:sector-resolvent}\\
  \calM^\lambda_n(\bK)
  &=\langle n,\bK;\lambda|\psi_e^\dagger({\bf 0})|0\rangle .
  \label{eq:sector-M}
\end{align}
Here $\calM$ denotes the exact overlap of the local electron with an exact charge-$e$ eigenstate.  When a sector is approximated by asymptotic scattering states, we use a lower-case contact amplitude $m_\lambda$ for the corresponding low-energy wave-function overlap.

The local electron operator probes the short-distance part of the multi-anyon wave function.  We first define the two-body contact exponents, since these are the building blocks used below.  For a pair of anyonic objects \(x,y\) with relative momentum \(p\), we write
\begin{equation}
  |m_{xy}(p)|^2\sim p^{2\gamma_{xy}},\qquad p\to0 .
  \label{eq:gamma-pair}
\end{equation}
For Abelian anyons this power is fixed by the short-distance statistical wave function.  If the mutual braiding phase of \(x\) and \(y\) is \(e^{2\pi\ii\alpha_{xy}}\), then a local source in relative angular-momentum channel \(\ell\) has
\begin{equation}
  \gamma_{xy}=|\ell-\alpha_{xy}|.
  \label{eq:gamma-xy}
\end{equation}
The allowed values of \(\ell\) are set by the microscopic tunneling matrix element and by the local symmetry channel of the electron operator.  The first observed onset is set by the sector with the lowest threshold energy.  If several thresholds overlap, the leading allowed local channel with the smallest contact exponent gives the strongest nonanalyticity.

For a compact \(e\) excitation there is no internal soft relative momentum, so \(\gamma_e=0\).  For the \(ba\) sector there is one soft relative momentum, and Eq.~\eqref{eq:gamma-pair} defines \(\gamma_{ba}\).  For a scalar local channel with \(\ell=0\), the minimal Laughlin choice \(b=a^2\) has \(\alpha_{ba}=2/3\) and gives \(\gamma_{ba}=2/3\).  We keep \(\gamma_{ba}\) explicit because lattice-scale tunneling matrix elements or selection rules can choose a different leading channel.  For the three-anyon sector \(aaa\), there are two independent Jacobi momenta, but at the clean activation threshold all pairwise relative momenta vanish with the same scale.  In the minimal Laughlin channel a symmetric representative of the contact factor is
\begin{equation}
  |m_{aaa}|^2\sim \prod_{i<j}|\bk_i-\bk_j|^{2\alpha},\qquad \gamma_{aaa}=3\alpha .
  \label{eq:maaa-product}
\end{equation}
For Laughlin \(\nu=1/3\), \(\alpha=\gamma_{aa}=1/3\), so \(\gamma_{aaa}=1\).

The clean local STM thresholds are fixed by phase-space counting.  For an \(N\)-body continuum with \(\Omega=\omega-\sum_i\Delta_i\),
\begin{equation}
  g_{0}(\omega)
  \sim
  \int \prod_{i=1}^{N}\dd^2k_i\,
  |m(\{\bk_i\})|^2
  \delta\!\left[\Omega-\sum_i{k_i^2\over 2M_i}\right].
\end{equation}
After rescaling \(\bk_i=\sqrt{2M_i\Omega}\,\tilde{\bk}_i\), the momentum integrals and the energy delta function contribute the phase-space factor \(\Omega^{N-1}\), and the contact matrix element supplies the additional factor \(\Omega^\gamma\).  Therefore
\begin{align}
  g_{0,aaa}(\omega)
  &\sim
  \Th(\omega-3\Delta_a)(\omega-3\Delta_a)^{2+\gamma_{aaa}},
  \label{eq:cleanaaa}\\
  g_{0,ba}(\omega)
  &\sim
  \Th[\omega-(\Delta_b+\Delta_a)]
  [\omega-(\Delta_b+\Delta_a)]^{1+\gamma_{ba}},
\end{align}
whereas a compact \(e\) excitation has no continuum phase space and gives
\(g_{0,e}(\omega)\sim \Th(\omega-\Delta_e)(\omega-\Delta_e)^0\).
These distinct threshold powers separate compact \(e\), partially bound \(ba\), and unbound \(aaa\) electron-addition sectors, thereby probing the binding hierarchy~\cite{Morampudi2017Threshold}.

Residual interactions renormalize the gaps, masses, and contact amplitudes in these formulas.  For short-range interactions, as long as the final state remains an unbound continuum with an isolated quadratic threshold, the powers are unchanged.  Unscreened Coulomb interactions can instead modify the asymptotic contact problem.  A metallic gate at distance \(d_g\) cuts off the long-range tail beyond \(d_g\), so the statistical contact exponent is recovered in the low-energy regime whose relative wavelength exceeds \(d_g\), parametrically \(k\sim\sqrt{2M\Omega}\lesssim d_g^{-1}\).  If interactions bind constituents, the relevant low-energy sector changes from \(aaa\) to \(ba\) or \(e\), and the corresponding threshold power should be used.

\begin{figure}[t]
\centering
\includegraphics[width=\linewidth]{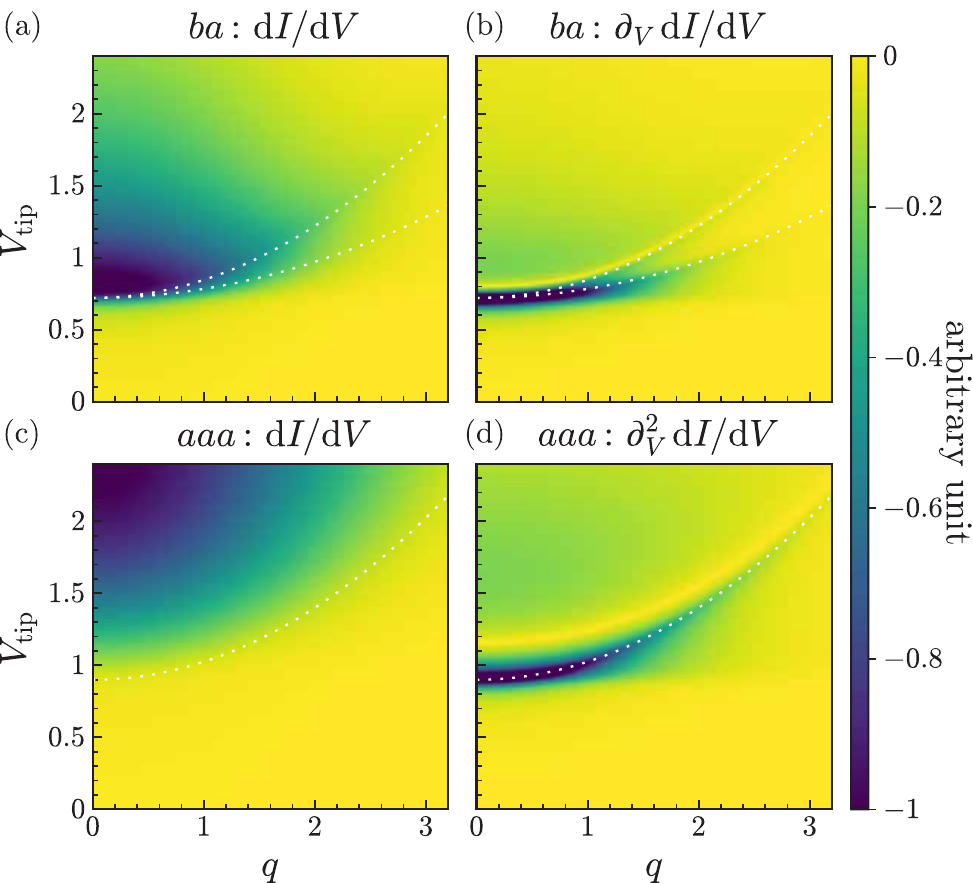}

\caption{Fourier-transform STS signals from the full numerical convolution kernel.  (a) $dI/dV$ for the $ba$ sector.  The two dashed guides correspond to scattering $b$ and scattering $a$.  (b) First energy derivative of the same $ba$ signal.  (c) $dI/dV$ for the $aaa$ sector.  (d) Second energy derivative of the $aaa$ signal.  The dashed guide is $\omega_{aaa}(q)=3\Delta_a+q^2/(8M_a)$.}
\label{fig:qpi}

\end{figure}

\begin{table}[t]
\caption{Clean STM onsets and fixed-$q$ Born-QPI branch singularities.  Smooth prefactors and step functions are suppressed.}
\label{tab:exponents}
\begingroup
\renewcommand{\arraystretch}{1.2}
\setlength{\tabcolsep}{0pt}
\begin{tabular*}{0.48\textwidth}{@{\extracolsep{\fill}}lll}
\hline\hline
Channel & Clean $dI/dV$ onset & Born-QPI singularity \\
\hline
compact $e$
& $(\omega-\Delta_e)^0$
& $[\omega_e(q)-\omega]^{-1/2}$ \\
$ba$, scatter $b$
& $[\omega-(\Delta_b+\Delta_a)]^{1+\gamma_{ba}}$
& $[\omega_{ba}^{(b)}(q)-\omega]^{1/2}$ \\
$ba$, scatter $a$
& $[\omega-(\Delta_b+\Delta_a)]^{1+\gamma_{ba}}$
& $[\omega_{ba}^{(a)}(q)-\omega]^{1/2}$ \\
$aaa$
& $[\omega-3\Delta_a]^{2+\gamma_{aaa}}$
& $[\omega_{aaa}(q)-\omega]^{3/2+\gamma_{aa}}$ \\
\hline\hline
\end{tabular*}
\endgroup
\end{table}

We now consider STS near a weak impurity as in Fig.~\ref{fig:stm}(a).
We define
\begin{equation}
  \rho_{\bq}
  =\int d^2{\bf r}\,e^{-i\bq\cdot{\bf r}}\rho({\bf r})
  =\sum_{\bk}c^\dagger_{\bk}c_{\bk+\bq}.
\end{equation}
A scalar impurity is therefore
\begin{equation}
  H_{\rm dis}
  =\int d^2{\bf r}\,U({\bf r})\rho({\bf r})
  ={1\over{\cal A}}\sum_{\bq}U_{\bq}\rho_{-\bq}.
\end{equation}
Here $\cal A$ is the sample area. The impurity modulates the local tunneling conductance, whose Fourier component is
\begin{equation}
  \delta g(\bq,\omega)
  =\int d^2{\bf r}\,e^{-i\bq\cdot{\bf r}}\delta g({\bf r},\omega)
  =-{1\over\pi}\Imop\delta G_{\rm diag}^R(\bq,\omega).
\end{equation}
In first Born approximation, let \(|n,\bP\rangle\) and \(|m,\bP\rangle\)
denote exact clean charge-\(e\) eigenstates.  The same exact electron overlap
in Eq.~\eqref{eq:sector-M} is now written as
\begin{align}
  \calM_n(\bP)
  &=\langle n,\bP|\psi_e^\dagger({\bf 0})|0\rangle,
  \nonumber\\
  \rho_{nm}(\bP+\bq,\bP)
  &=\langle n,\bP+\bq|\rho_{-\bq}|m,\bP\rangle .
  \label{eq:M-rho-def}
\end{align}
Then
\begin{align}
\delta G_{\rm diag}^R(\bq,\omega)
&=U_{\bq}\sum_{\bP,n,m}
{\calM_n^*(\bP+\bq)
\rho_{nm}(\bP+\bq,\bP)
\over
\omega-E_n(\bP+\bq)+\ii0^+}
\nonumber\\
&\quad\times
{\calM_m(\bP)\over \omega-E_m(\bP)+\ii0^+}.
\label{eq:born}
\end{align}
Thus the \(\bq\) component of QPI probes
\(\calM^*G_0^R\rho_{-\bq}G_0^R\calM\), not a final-state density of states.

For any parabolic object in Eq.~\eqref{eq:generic-dispersion}, the retarded--retarded bubble is
\begin{equation}
  \Pi_s(q,\omega)=
  \int\dd^2k\,
  {1\over\omega-E_s(\bk+\bq)+\ii0^+}
  {1\over\omega-E_s(\bk)+\ii0^+}.
\end{equation}
Its branch point occurs when the same object is on shell before and after absorbing momentum \(\bq\).  For a parabolic band this occurs at \(\bk=-\bq/2\) and \(\bk+\bq=\bq/2\), giving
\begin{equation}
  \omega_s(q)=\Delta_s+{q^2\over8M_s}.
\end{equation}
The corresponding weak-scattering singularity is~\cite{WangLee2003,Capriotti2003,PeregBarneaFranz2003,SimonReview,BenaReview}
\begin{equation}
  -{1\over\pi}\Imop\Pi_s(q,\omega)
  \sim
  [\omega_s(q)-\omega]^{-1/2} .
  \label{eq:single-bubble}
\end{equation}
Here and below the QPI branch nonanalyticities are understood on the side $\omega<\omega_{\rm branch}(q)$, with smooth activation and support factors suppressed.

Now consider the $ba$ sector and let disorder scatter $b$.  In a free-continuum representation, the same low-energy electron overlap appears as a plane-wave contact amplitude $m_{ba}(\bk_b,\bk_a)$.  It is convenient to write
\begin{equation}
  m_{ba}^*(\bk_b+\bq,\bk_a)
  m_{ba}(\bk_b,\bk_a)
  =
  W_b(q,\bk_b)W_a(q,\bk_a),
  \label{eq:factorization}
\end{equation}
which defines smooth weights $W_{a,b}$ on the branch.  This factorized form is only a convenient notation. The physical assumption is that the matrix element is smooth at fixed nonzero \(q\).  At the branch where \(b\) is scattered, the singular configuration has
\(\bk_b\simeq-\bq/2\), \(\bk_b+\bq\simeq\bq/2\), and the spectator \(a\) near its minimum.  Hence the \(b-a\) relative momentum is \(O(q)\).  In the fixed-\(q\) limit the small parameter is the energy detuning \(\delta_b(q,\omega)=\omega_{ba}^{(b)}(q)-\omega\to0\).  The contact factor \(p_{ba}^{2\gamma_{ba}}\) is therefore a smooth envelope, typically \(q^{2\gamma_{ba}}\), rather than a power of \(\delta_b\).  The fixed-\(q\) exponent comes only from the \(b\) retarded--retarded bubble and the spectator \(a\) density of states.  Matrix-element selection rules can suppress the intensity, or increase the apparent power if the leading smooth envelope vanishes, but they do not move the on-shell branch.

With the spectator kinetic energy $\eps_a=k_a^2/(2M_a)$, the remaining $b$ integral is the reduced-energy bubble
\begin{align}
  \Pi_b(q,E)
  &=\int \dd^2k_b\,
  { W_b(q,\bk_b)\over E-E_b(\bk_b+\bq)+\ii0^+ }
  \nonumber\\
  &\quad\times
  {1\over E-E_b(\bk_b)+\ii0^+} .
  \label{eq:Pi-b-main}
\end{align}
The singular part can be written as
\begin{align}
  \delta g_{ba,b}(q,\omega)
  &\propto
  \int_0^\infty\dd\eps_a\,\rho_a(\eps_a;q)
  \nonumber\\
  &\quad\times
  \left[-{1\over\pi}\Imop\Pi_b(q,\omega-\Delta_a-\eps_a)\right],
  \label{eq:convbamain}
\end{align}
where
\begin{equation}
  \rho_a(\eps_a;q)=
  \int\dd^2k_a\,W_a(q,\bk_a)
  \delta\!\left(\eps_a-{k_a^2\over2M_a}\right).
\end{equation}
For smooth $W_a$, $\rho_a\sim \eps_a^0$.  Eq.~\eqref{eq:single-bubble} and \eqref{eq:convbamain} give
\begin{equation}
  \omega_{ba}^{(b)}(q)=\Delta_b+\Delta_a+{q^2\over8M_b},
  \qquad
  \delta g_{ba,b}\sim [\omega_{ba}^{(b)}(q)-\omega]^{1/2}.
\end{equation}
Scattering the $a$ anyon instead gives
\begin{align}
  \omega_{ba}^{(a)}(q)&=\Delta_b+\Delta_a+{q^2\over8M_a},
  \nonumber\\
  \delta g_{ba,a}&\sim
  [\omega_{ba}^{(a)}(q)-\omega]^{1/2}.
\end{align}

For the three-anyon branch, disorder scatters one $a$ and the other two are spectators.  Their local spectator density obeys
\begin{equation}
  \rho_{aa}(\eps_{aa};q)
  \sim \eps_{aa}^{1+\gamma_{aa}},
\end{equation}
where one power comes from two-particle two-dimensional phase space and $\gamma_{aa}=\alpha$ is the soft $a-a$ contact exponent.  Hence
\begin{equation}
  \delta g_{aaa}(q,\omega)
  \propto
  \int_0^\Lambda\dd\eps_{aa}
  {\eps_{aa}^{1+\gamma_{aa}}
  \over
  \sqrt{\omega_{aaa}(q)-\omega+\eps_{aa}}}.
  \label{eq:convaaamain}
\end{equation}
The branch and singularity are
\begin{equation}
  \omega_{aaa}(q)=3\Delta_a+{q^2\over8M_a},
\end{equation}
\begin{equation}
  \delta g_{aaa}
  \sim
  [\omega_{aaa}(q)-\omega]^{3/2+\gamma_{aa}} .
\end{equation}
For Laughlin $\nu=1/3$, $\gamma_{aa}=\alpha=1/3$, giving exponent $11/6$.
\begin{figure}[t]
\centering
\includegraphics[width=\linewidth]{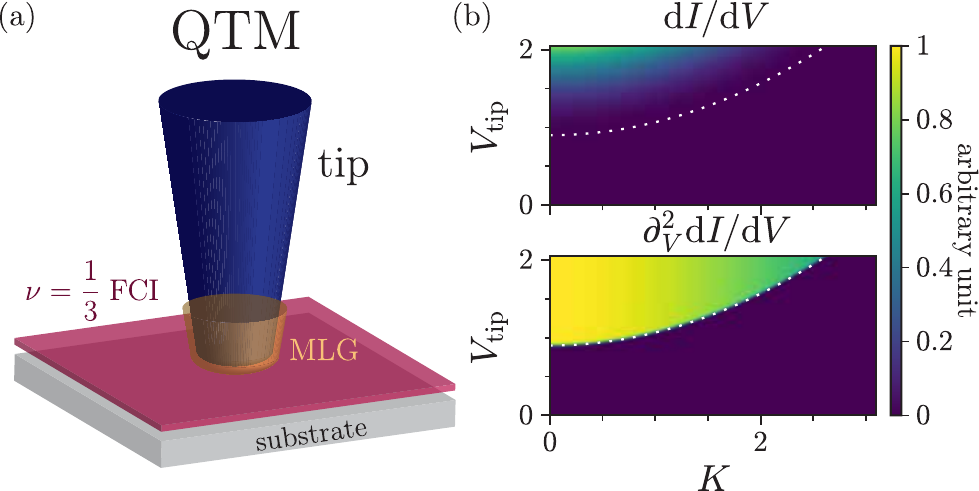}

\caption{QTM.  (a) A monolayer graphene (MLG) QTM probe forms a planar tunnel junction with the FCI sample.  The twist angle selects the injected electron momentum $\bK$.  (b) $dI/dV$ and $\partial_V^2(\dd I/\dd V)$ for the $aaa$ contribution to $A^+_{e,{\rm FCI}}(\bK,eV)$ in the small-pocket, elastic-tunneling limit.}
\label{fig:qtm}

\end{figure}

\emph{QTM.---}
Quantum twisting microscopy (QTM) replaces the local STM point contact by a planar van der Waals tunnel junction. Because tunneling is coherent over many unit cells, in-plane momentum is conserved up to reciprocal lattice vectors. In a given Umklapp channel, an electron with momentum \(\bp\) in the monolayer graphene probe tunnels into the FCI sample at momentum \(\bp+\bK(\theta)\), where \(\bK(\theta)\) is the twist-dependent mismatch between the rotated graphene Brillouin zone and the sample Brillouin zone. Varying the twist angle moves this selected momentum through the sample Brillouin zone.

At low temperature,
\begin{equation}
  {\dd I\over\dd V}
  \propto
  \int_{\rm MLG\,FS}\dd\ell_{\bp}\,
  |t(\bp,\bK)|^2
  A^+_{e,{\rm FCI}}(\bp+\bK,eV).
  \label{eq:qtm-tipfs}
\end{equation}
Here $A^+_{e,{\rm FCI}}$ denotes the electron-addition spectral function, and smooth graphene spectral and velocity factors have been absorbed into \(t\).  For a small graphene pocket, Eq.~\eqref{eq:qtm-tipfs} reduces to \(dI/dV\propto A^+_{e,{\rm FCI}}(\bK,eV)\), with finite pocket size and temperature only broadening the threshold when they are small compared with the FCI edge scale.  Thus QTM is a momentum-resolved version of threshold spectroscopy~\cite{Morampudi2017Threshold}: the edge power and its \(K\)-dependence encode how the injected electron fractionalizes. This is not a retarded--retarded impurity bubble: QTM measures the spectral function itself, while STS-QPI measures the Born kernel \(\calM^*G_0^R\rho_{-\bq}G_0^R\calM\).

The QTM exponents are obtained by fixing the total momentum of the electron-addition final state.  For the compact and two-body sectors,
\begin{align}
  E_e^{\rm QTM}(K)&=\Delta_e+{K^2\over2M_e},\nonumber\\
  E_{ba}^{\rm QTM}(K)&=\Delta_b+\Delta_a+{K^2\over2(M_b+M_a)}.
\end{align}
Writing the relative momentum as \(\bp\), the fixed-\(K\) spectral
weight near this edge is
\begin{align}
  A_{ba}(\bK,\omega)
  &\sim
  \int \dd^2p\,
  |m_{ba}(p)|^2
  \delta\!\left[
  \omega-E_{ba}^{\rm QTM}(K)-{p^2\over2\mu_{ba}}
  \right],
  \label{eq:qtm-ba-integral}
\end{align}
where \(\mu_{ba}=M_bM_a/(M_b+M_a)\).  Since
\(|m_{ba}(p)|^2\sim p^{2\gamma_{ba}}\), the two-dimensional relative
phase space gives
\begin{equation}
  A_{ba}(\bK,\omega)
  \sim
  [\omega-E_{ba}^{\rm QTM}(K)]^{\gamma_{ba}}
  \Th[\omega-E_{ba}^{\rm QTM}(K)] .
  \label{eq:qtm-ba-power}
\end{equation}

For the unbound three-anyon sector,
\begin{equation}
  E_{aaa}^{\rm QTM}(K)=3\Delta_a+{K^2\over6M_a}.
  \label{eq:metal-fci-edge}
\end{equation}
At fixed total momentum, the three-body continuum has two independent relative momenta.  The relative phase space gives one power of the excess energy, and the contact factor contributes \(\gamma_{aaa}\), giving
\begin{equation}
  A_{aaa}(\bK,\omega)
  \sim
  [\omega-E_{aaa}^{\rm QTM}(K)]^{1+\gamma_{aaa}}
  \Th[\omega-E_{aaa}^{\rm QTM}(K)] .
  \label{eq:qtm-aaa-power}
\end{equation}
For Laughlin \(\nu=1/3\), \(1+\gamma_{aaa}=2\), so the raw
\(dI/dV\) threshold is smooth.  Fig.~\ref{fig:qtm}(b) therefore shows
both \(dI/dV\) and \(\partial_V^2(\dd I/\dd V)\) for the same \(aaa\) spectrum.

The key contrast is the momentum constraint.  For the \(aaa\) sector,
\begin{equation}
  \omega_{aaa}^{\rm QPI}(q)=3\Delta_a+{q^2\over8M_a},
  \qquad
  E_{aaa}^{\rm QTM}(K)=3\Delta_a+{K^2\over6M_a}.
\end{equation}
For a general anyon Bloch band, the same distinction remains: the QPI
branch is obtained from equal-energy scattering of one constituent,
whereas the QTM edge is obtained by minimizing the sum of constituent
Bloch energies at fixed total momentum.  Isolated quadratic minima give
the powers above, whereas saddles, flat directions, or higher-order minima can
modify the threshold exponents.

\begin{table}[t]
\caption{QTM fixed-$K$ spectral features.  Smooth prefactors and step functions are suppressed.}
\label{tab:qtm-exponents}
\begin{ruledtabular}
\begin{tabular}{ll}
Channel & QTM feature \\
\hline
compact $e$
& $\delta[\omega-E_e^{\rm QTM}(K)]$ \\
$ba$
& $[\omega-E_{ba}^{\rm QTM}(K)]^{\gamma_{ba}}$ \\
$aaa$
& $[\omega-E_{aaa}^{\rm QTM}(K)]^{1+\gamma_{aaa}}$ \\
\end{tabular}
\end{ruledtabular}
\end{table}

\emph{Discussion.---}
The proposed measurements are subject to the usual matrix-element constraints of tunneling spectroscopy.  In FQH STS, the angular-momentum structure of the tunneling electron can strongly affect LDOS peak weights even when excitation energies are unchanged~\cite{PuCompositeFermionSTM}.  In our notation, the same physics appears in the contact exponents \(\gamma_\lambda\) for clean activation and in the smooth QPI envelopes \(W_s(q,\bk)\).  If the leading local channel is allowed, these factors control intensity but not branch positions; if it is suppressed, the visible signal can be weak and the apparent power can be raised by the first allowed channel.

This matrix-element issue motivates using conductance derivatives.  For Laughlin \(\nu=1/3\), the \(aaa\) QPI branch has \(\delta g_{aaa}\sim[\omega_{aaa}(q)-\omega]^{11/6}\), so it need not appear as a sharp peak in raw \(dI/dV\).  The second derivative exposes the weak singularity in Figs.~\ref{fig:stm} and \ref{fig:qpi}.  The clean activation exponents are independently useful because they distinguish compact \(e\), \(ba\), and \(aaa\) sectors, thereby testing the binding hierarchy.

The parabolic formulas are local expansions of anyon Bloch bands near isolated minima.  For a general FCI anyon band, QPI extracts equal-energy scattering of one constituent, while QTM extracts the continuum edge at fixed total momentum.  Combining them provides a consistency check on the extracted mass and a probe of the binding hierarchy that controls dilute anyon gases in moir\'e FCIs.

Finally, the same probes may help distinguish an anyon-driven superconductor from a conventional weak-coupling BCS scenario. In the weak-coupling BCS case, one naturally expects an approximately particle-hole symmetric tunneling spectrum near zero bias, whereas in an anyon-driven superconductor there is no reason for the positive- and negative-bias spectra to be symmetric. In the proximate normal state above \(T_c\), this anyon-based scenario may retain activated electron tunneling with the fractionalized thresholds summarized in Tables~\ref{tab:exponents} and \ref{tab:qtm-exponents}, together with QPI or QTM dispersions tied to constituent anyons rather than compact electron-like quasiparticles. Particle-hole asymmetry by itself, however, is not unique to anyon physics, since unrelated strong interactions or disorder can also produce it. The more specific diagnostic is the combination of bias asymmetry with activation exponents.

\textbf{Acknowledgments.}
We thank Zhengyan Darius Shi and Michael P. Zaletel for helpful discussions. T.W. is grateful for support from the Harvard Quantum Initiative Fellowship and the Simons Collaboration on Ultra-Quantum Matter, which is a grant from the Simons Foundation (Grant No. 651440). T.S. was supported by the U.S. Department of Energy under Grant DE-SC0008739.

\bibliographystyle{apsrev4-2}
\bibliography{paper}

\begin{thebibliography}{54}%
\makeatletter
\providecommand \@ifxundefined [1]{%
 \@ifx{#1\undefined}
}%
\providecommand \@ifnum [1]{%
 \ifnum #1\expandafter \@firstoftwo
 \else \expandafter \@secondoftwo
 \fi
}%
\providecommand \@ifx [1]{%
 \ifx #1\expandafter \@firstoftwo
 \else \expandafter \@secondoftwo
 \fi
}%
\providecommand \natexlab [1]{#1}%
\providecommand \enquote  [1]{``#1''}%
\providecommand \bibnamefont  [1]{#1}%
\providecommand \bibfnamefont [1]{#1}%
\providecommand \citenamefont [1]{#1}%
\providecommand \href@noop [0]{\@secondoftwo}%
\providecommand \href [0]{\begingroup \@sanitize@url \@href}%
\providecommand \@href[1]{\@@startlink{#1}\@@href}%
\providecommand \@@href[1]{\endgroup#1\@@endlink}%
\providecommand \@sanitize@url [0]{\catcode `\\12\catcode `\$12\catcode `\&12\catcode `\#12\catcode `\^12\catcode `\_12\catcode `\%12\relax}%
\providecommand \@@startlink[1]{}%
\providecommand \@@endlink[0]{}%
\providecommand \url  [0]{\begingroup\@sanitize@url \@url }%
\providecommand \@url [1]{\endgroup\@href {#1}{\urlprefix }}%
\providecommand \urlprefix  [0]{URL }%
\providecommand \Eprint [0]{\href }%
\providecommand \doibase [0]{https://doi.org/}%
\providecommand \selectlanguage [0]{\@gobble}%
\providecommand \bibinfo  [0]{\@secondoftwo}%
\providecommand \bibfield  [0]{\@secondoftwo}%
\providecommand \translation [1]{[#1]}%
\providecommand \BibitemOpen [0]{}%
\providecommand \bibitemStop [0]{}%
\providecommand \bibitemNoStop [0]{.\EOS\space}%
\providecommand \EOS [0]{\spacefactor3000\relax}%
\providecommand \BibitemShut  [1]{\csname bibitem#1\endcsname}%
\let\auto@bib@innerbib\@empty
\bibitem [{\citenamefont {Laughlin}(1983)}]{Laughlin1983}%
  \BibitemOpen
  \bibfield  {author} {\bibinfo {author} {\bibfnamefont {R.~B.}\ \bibnamefont {Laughlin}},\ }\href {https://doi.org/10.1103/PhysRevLett.50.1395} {\bibfield  {journal} {\bibinfo  {journal} {Phys. Rev. Lett.}\ }\textbf {\bibinfo {volume} {50}},\ \bibinfo {pages} {1395} (\bibinfo {year} {1983})}\BibitemShut {NoStop}%
\bibitem [{\citenamefont {Arovas}\ \emph {et~al.}(1984)\citenamefont {Arovas}, \citenamefont {Schrieffer},\ and\ \citenamefont {Wilczek}}]{Arovas1984}%
  \BibitemOpen
  \bibfield  {author} {\bibinfo {author} {\bibfnamefont {D.}~\bibnamefont {Arovas}}, \bibinfo {author} {\bibfnamefont {J.~R.}\ \bibnamefont {Schrieffer}},\ and\ \bibinfo {author} {\bibfnamefont {F.}~\bibnamefont {Wilczek}},\ }\href {https://doi.org/10.1103/PhysRevLett.53.722} {\bibfield  {journal} {\bibinfo  {journal} {Phys. Rev. Lett.}\ }\textbf {\bibinfo {volume} {53}},\ \bibinfo {pages} {722} (\bibinfo {year} {1984})}\BibitemShut {NoStop}%
\bibitem [{\citenamefont {Zeng}\ \emph {et~al.}(2023)\citenamefont {Zeng}, \citenamefont {Xia}, \citenamefont {Kang}, \citenamefont {Zhu}, \citenamefont {Kn{\"u}ppel}, \citenamefont {Vaswani}, \citenamefont {Watanabe}, \citenamefont {Taniguchi}, \citenamefont {Mak},\ and\ \citenamefont {Shan}}]{ZengMoTe2}%
  \BibitemOpen
  \bibfield  {author} {\bibinfo {author} {\bibfnamefont {Y.}~\bibnamefont {Zeng}}, \bibinfo {author} {\bibfnamefont {Z.}~\bibnamefont {Xia}}, \bibinfo {author} {\bibfnamefont {K.}~\bibnamefont {Kang}}, \bibinfo {author} {\bibfnamefont {J.}~\bibnamefont {Zhu}}, \bibinfo {author} {\bibfnamefont {P.}~\bibnamefont {Kn{\"u}ppel}}, \bibinfo {author} {\bibfnamefont {C.}~\bibnamefont {Vaswani}}, \bibinfo {author} {\bibfnamefont {K.}~\bibnamefont {Watanabe}}, \bibinfo {author} {\bibfnamefont {T.}~\bibnamefont {Taniguchi}}, \bibinfo {author} {\bibfnamefont {K.~F.}\ \bibnamefont {Mak}},\ and\ \bibinfo {author} {\bibfnamefont {J.}~\bibnamefont {Shan}},\ }\href {https://doi.org/10.1038/s41586-023-06452-3} {\bibfield  {journal} {\bibinfo  {journal} {Nature}\ }\textbf {\bibinfo {volume} {622}},\ \bibinfo {pages} {69} (\bibinfo {year} {2023})}\BibitemShut {NoStop}%
\bibitem [{\citenamefont {Park}\ \emph {et~al.}(2023)\citenamefont {Park}, \citenamefont {Cai}, \citenamefont {Anderson}, \citenamefont {Zhang}, \citenamefont {Zhu}, \citenamefont {Liu}, \citenamefont {Wang}, \citenamefont {Holtzmann}, \citenamefont {Hu}, \citenamefont {Liu}, \citenamefont {Taniguchi}, \citenamefont {Watanabe}, \citenamefont {Chu}, \citenamefont {Cao}, \citenamefont {Fu}, \citenamefont {Yao}, \citenamefont {Chang}, \citenamefont {Cobden}, \citenamefont {Xiao},\ and\ \citenamefont {Xu}}]{ParkMoTe2}%
  \BibitemOpen
  \bibfield  {author} {\bibinfo {author} {\bibfnamefont {H.}~\bibnamefont {Park}}, \bibinfo {author} {\bibfnamefont {J.}~\bibnamefont {Cai}}, \bibinfo {author} {\bibfnamefont {E.}~\bibnamefont {Anderson}}, \bibinfo {author} {\bibfnamefont {Y.}~\bibnamefont {Zhang}}, \bibinfo {author} {\bibfnamefont {J.}~\bibnamefont {Zhu}}, \bibinfo {author} {\bibfnamefont {X.}~\bibnamefont {Liu}}, \bibinfo {author} {\bibfnamefont {C.}~\bibnamefont {Wang}}, \bibinfo {author} {\bibfnamefont {W.}~\bibnamefont {Holtzmann}}, \bibinfo {author} {\bibfnamefont {C.}~\bibnamefont {Hu}}, \bibinfo {author} {\bibfnamefont {Z.}~\bibnamefont {Liu}}, \bibinfo {author} {\bibfnamefont {T.}~\bibnamefont {Taniguchi}}, \bibinfo {author} {\bibfnamefont {K.}~\bibnamefont {Watanabe}}, \bibinfo {author} {\bibfnamefont {J.-H.}\ \bibnamefont {Chu}}, \bibinfo {author} {\bibfnamefont {T.}~\bibnamefont {Cao}}, \bibinfo {author} {\bibfnamefont {L.}~\bibnamefont {Fu}}, \bibinfo {author} {\bibfnamefont {W.}~\bibnamefont {Yao}}, \bibinfo {author}
  {\bibfnamefont {C.-Z.}\ \bibnamefont {Chang}}, \bibinfo {author} {\bibfnamefont {D.}~\bibnamefont {Cobden}}, \bibinfo {author} {\bibfnamefont {D.}~\bibnamefont {Xiao}},\ and\ \bibinfo {author} {\bibfnamefont {X.}~\bibnamefont {Xu}},\ }\href {https://doi.org/10.1038/s41586-023-06536-0} {\bibfield  {journal} {\bibinfo  {journal} {Nature}\ }\textbf {\bibinfo {volume} {622}},\ \bibinfo {pages} {74} (\bibinfo {year} {2023})},\ \Eprint {https://arxiv.org/abs/2308.02657} {arXiv:2308.02657} \BibitemShut {NoStop}%
\bibitem [{\citenamefont {Lu}\ \emph {et~al.}(2024)\citenamefont {Lu}, \citenamefont {Han}, \citenamefont {Yao}, \citenamefont {Reddy}, \citenamefont {Yang}, \citenamefont {Seo}, \citenamefont {Watanabe}, \citenamefont {Taniguchi}, \citenamefont {Fu},\ and\ \citenamefont {Ju}}]{LuMLG}%
  \BibitemOpen
  \bibfield  {author} {\bibinfo {author} {\bibfnamefont {Z.}~\bibnamefont {Lu}}, \bibinfo {author} {\bibfnamefont {T.}~\bibnamefont {Han}}, \bibinfo {author} {\bibfnamefont {Y.}~\bibnamefont {Yao}}, \bibinfo {author} {\bibfnamefont {A.~P.}\ \bibnamefont {Reddy}}, \bibinfo {author} {\bibfnamefont {J.}~\bibnamefont {Yang}}, \bibinfo {author} {\bibfnamefont {J.}~\bibnamefont {Seo}}, \bibinfo {author} {\bibfnamefont {K.}~\bibnamefont {Watanabe}}, \bibinfo {author} {\bibfnamefont {T.}~\bibnamefont {Taniguchi}}, \bibinfo {author} {\bibfnamefont {L.}~\bibnamefont {Fu}},\ and\ \bibinfo {author} {\bibfnamefont {L.}~\bibnamefont {Ju}},\ }\href {https://doi.org/10.1038/s41586-023-07010-7} {\bibfield  {journal} {\bibinfo  {journal} {Nature}\ }\textbf {\bibinfo {volume} {626}},\ \bibinfo {pages} {759} (\bibinfo {year} {2024})}\BibitemShut {NoStop}%
\bibitem [{\citenamefont {Lu}\ \emph {et~al.}(2025{\natexlab{a}})\citenamefont {Lu}, \citenamefont {Han}, \citenamefont {Yao}, \citenamefont {Hadjri}, \citenamefont {Yang}, \citenamefont {Seo}, \citenamefont {Shi}, \citenamefont {Ye}, \citenamefont {Watanabe}, \citenamefont {Taniguchi},\ and\ \citenamefont {Ju}}]{LuExtendedQAH2025}%
  \BibitemOpen
  \bibfield  {author} {\bibinfo {author} {\bibfnamefont {Z.}~\bibnamefont {Lu}}, \bibinfo {author} {\bibfnamefont {T.}~\bibnamefont {Han}}, \bibinfo {author} {\bibfnamefont {Y.}~\bibnamefont {Yao}}, \bibinfo {author} {\bibfnamefont {Z.}~\bibnamefont {Hadjri}}, \bibinfo {author} {\bibfnamefont {J.}~\bibnamefont {Yang}}, \bibinfo {author} {\bibfnamefont {J.}~\bibnamefont {Seo}}, \bibinfo {author} {\bibfnamefont {L.}~\bibnamefont {Shi}}, \bibinfo {author} {\bibfnamefont {S.}~\bibnamefont {Ye}}, \bibinfo {author} {\bibfnamefont {K.}~\bibnamefont {Watanabe}}, \bibinfo {author} {\bibfnamefont {T.}~\bibnamefont {Taniguchi}},\ and\ \bibinfo {author} {\bibfnamefont {L.}~\bibnamefont {Ju}},\ }\href {https://doi.org/10.1038/s41586-024-08470-1} {\bibfield  {journal} {\bibinfo  {journal} {Nature}\ }\textbf {\bibinfo {volume} {637}},\ \bibinfo {pages} {1090} (\bibinfo {year} {2025}{\natexlab{a}})}\BibitemShut {NoStop}%
\bibitem [{\citenamefont {Aronson}\ \emph {et~al.}(2025)\citenamefont {Aronson}, \citenamefont {Han}, \citenamefont {Lu}, \citenamefont {Yao}, \citenamefont {Butler}, \citenamefont {Watanabe}, \citenamefont {Taniguchi}, \citenamefont {Ju},\ and\ \citenamefont {Ashoori}}]{AronsonFCI2025}%
  \BibitemOpen
  \bibfield  {author} {\bibinfo {author} {\bibfnamefont {S.~H.}\ \bibnamefont {Aronson}}, \bibinfo {author} {\bibfnamefont {T.}~\bibnamefont {Han}}, \bibinfo {author} {\bibfnamefont {Z.}~\bibnamefont {Lu}}, \bibinfo {author} {\bibfnamefont {Y.}~\bibnamefont {Yao}}, \bibinfo {author} {\bibfnamefont {J.~P.}\ \bibnamefont {Butler}}, \bibinfo {author} {\bibfnamefont {K.}~\bibnamefont {Watanabe}}, \bibinfo {author} {\bibfnamefont {T.}~\bibnamefont {Taniguchi}}, \bibinfo {author} {\bibfnamefont {L.}~\bibnamefont {Ju}},\ and\ \bibinfo {author} {\bibfnamefont {R.~C.}\ \bibnamefont {Ashoori}},\ }\href {https://doi.org/10.1103/75gl-jzl6} {\bibfield  {journal} {\bibinfo  {journal} {Phys. Rev. X}\ }\textbf {\bibinfo {volume} {15}},\ \bibinfo {pages} {031026} (\bibinfo {year} {2025})}\BibitemShut {NoStop}%
\bibitem [{\citenamefont {Waters}\ \emph {et~al.}(2025)\citenamefont {Waters}, \citenamefont {Okounkova}, \citenamefont {Su}, \citenamefont {Zhou}, \citenamefont {Yao}, \citenamefont {Watanabe}, \citenamefont {Taniguchi}, \citenamefont {Xu}, \citenamefont {Zhang}, \citenamefont {Folk},\ and\ \citenamefont {Yankowitz}}]{WatersPentalayer}%
  \BibitemOpen
  \bibfield  {author} {\bibinfo {author} {\bibfnamefont {D.}~\bibnamefont {Waters}}, \bibinfo {author} {\bibfnamefont {A.}~\bibnamefont {Okounkova}}, \bibinfo {author} {\bibfnamefont {R.}~\bibnamefont {Su}}, \bibinfo {author} {\bibfnamefont {B.}~\bibnamefont {Zhou}}, \bibinfo {author} {\bibfnamefont {J.}~\bibnamefont {Yao}}, \bibinfo {author} {\bibfnamefont {K.}~\bibnamefont {Watanabe}}, \bibinfo {author} {\bibfnamefont {T.}~\bibnamefont {Taniguchi}}, \bibinfo {author} {\bibfnamefont {X.}~\bibnamefont {Xu}}, \bibinfo {author} {\bibfnamefont {Y.-H.}\ \bibnamefont {Zhang}}, \bibinfo {author} {\bibfnamefont {J.}~\bibnamefont {Folk}},\ and\ \bibinfo {author} {\bibfnamefont {M.}~\bibnamefont {Yankowitz}},\ }\href {https://doi.org/10.1103/PhysRevX.15.011045} {\bibfield  {journal} {\bibinfo  {journal} {Phys. Rev. X}\ }\textbf {\bibinfo {volume} {15}},\ \bibinfo {pages} {011045} (\bibinfo {year} {2025})},\ \Eprint {https://arxiv.org/abs/2408.10133} {arXiv:2408.10133} \BibitemShut {NoStop}%
\bibitem [{\citenamefont {Spanton}\ \emph {et~al.}(2018)\citenamefont {Spanton}, \citenamefont {Zibrov}, \citenamefont {Zhou}, \citenamefont {Taniguchi}, \citenamefont {Watanabe}, \citenamefont {Zaletel},\ and\ \citenamefont {Young}}]{Spanton2018FCI}%
  \BibitemOpen
  \bibfield  {author} {\bibinfo {author} {\bibfnamefont {E.~M.}\ \bibnamefont {Spanton}}, \bibinfo {author} {\bibfnamefont {A.~A.}\ \bibnamefont {Zibrov}}, \bibinfo {author} {\bibfnamefont {H.}~\bibnamefont {Zhou}}, \bibinfo {author} {\bibfnamefont {T.}~\bibnamefont {Taniguchi}}, \bibinfo {author} {\bibfnamefont {K.}~\bibnamefont {Watanabe}}, \bibinfo {author} {\bibfnamefont {M.~P.}\ \bibnamefont {Zaletel}},\ and\ \bibinfo {author} {\bibfnamefont {A.~F.}\ \bibnamefont {Young}},\ }\href {https://doi.org/10.1126/science.aan8458} {\bibfield  {journal} {\bibinfo  {journal} {Science}\ }\textbf {\bibinfo {volume} {360}},\ \bibinfo {pages} {62} (\bibinfo {year} {2018})},\ \Eprint {https://arxiv.org/abs/1706.06116} {arXiv:1706.06116 [cond-mat.str-el]} \BibitemShut {NoStop}%
\bibitem [{\citenamefont {Tsui}\ \emph {et~al.}(2025)\citenamefont {Tsui}, \citenamefont {He}, \citenamefont {Wang}, \citenamefont {Watanabe}, \citenamefont {Taniguchi}, \citenamefont {Zaletel},\ and\ \citenamefont {Yazdani}}]{TsuiHofstadterUnpublished}%
  \BibitemOpen
  \bibfield  {author} {\bibinfo {author} {\bibfnamefont {Y.-C.}\ \bibnamefont {Tsui}}, \bibinfo {author} {\bibfnamefont {M.}~\bibnamefont {He}}, \bibinfo {author} {\bibfnamefont {T.}~\bibnamefont {Wang}}, \bibinfo {author} {\bibfnamefont {K.}~\bibnamefont {Watanabe}}, \bibinfo {author} {\bibfnamefont {T.}~\bibnamefont {Taniguchi}}, \bibinfo {author} {\bibfnamefont {M.~P.}\ \bibnamefont {Zaletel}},\ and\ \bibinfo {author} {\bibfnamefont {A.}~\bibnamefont {Yazdani}}} (\bibinfo {year} {2025}),\ \bibinfo {note} {aPS Global Physics Summit 2025, contributed talk}\BibitemShut {NoStop}%
\bibitem [{\citenamefont {Xie}\ \emph {et~al.}(2021)\citenamefont {Xie}, \citenamefont {Pierce}, \citenamefont {Park}, \citenamefont {Parker}, \citenamefont {Khalaf}, \citenamefont {Ledwith}, \citenamefont {Cao}, \citenamefont {Lee}, \citenamefont {Chen}, \citenamefont {Forrester}, \citenamefont {Watanabe}, \citenamefont {Taniguchi}, \citenamefont {Vishwanath}, \citenamefont {Jarillo-Herrero},\ and\ \citenamefont {Yacoby}}]{XieMATBG}%
  \BibitemOpen
  \bibfield  {author} {\bibinfo {author} {\bibfnamefont {Y.}~\bibnamefont {Xie}}, \bibinfo {author} {\bibfnamefont {A.~T.}\ \bibnamefont {Pierce}}, \bibinfo {author} {\bibfnamefont {J.~M.}\ \bibnamefont {Park}}, \bibinfo {author} {\bibfnamefont {D.~E.}\ \bibnamefont {Parker}}, \bibinfo {author} {\bibfnamefont {E.}~\bibnamefont {Khalaf}}, \bibinfo {author} {\bibfnamefont {P.}~\bibnamefont {Ledwith}}, \bibinfo {author} {\bibfnamefont {Y.}~\bibnamefont {Cao}}, \bibinfo {author} {\bibfnamefont {S.~H.}\ \bibnamefont {Lee}}, \bibinfo {author} {\bibfnamefont {S.}~\bibnamefont {Chen}}, \bibinfo {author} {\bibfnamefont {P.~R.}\ \bibnamefont {Forrester}}, \bibinfo {author} {\bibfnamefont {K.}~\bibnamefont {Watanabe}}, \bibinfo {author} {\bibfnamefont {T.}~\bibnamefont {Taniguchi}}, \bibinfo {author} {\bibfnamefont {A.}~\bibnamefont {Vishwanath}}, \bibinfo {author} {\bibfnamefont {P.}~\bibnamefont {Jarillo-Herrero}},\ and\ \bibinfo {author} {\bibfnamefont {A.}~\bibnamefont {Yacoby}},\ }\href
  {https://doi.org/10.1038/s41586-021-04002-3} {\bibfield  {journal} {\bibinfo  {journal} {Nature}\ }\textbf {\bibinfo {volume} {600}},\ \bibinfo {pages} {439} (\bibinfo {year} {2021})},\ \Eprint {https://arxiv.org/abs/2107.10854} {arXiv:2107.10854} \BibitemShut {NoStop}%
\bibitem [{\citenamefont {Shi}\ and\ \citenamefont {Senthil}(2025{\natexlab{a}})}]{Shi2024_doping}%
  \BibitemOpen
  \bibfield  {author} {\bibinfo {author} {\bibfnamefont {Z.~D.}\ \bibnamefont {Shi}}\ and\ \bibinfo {author} {\bibfnamefont {T.}~\bibnamefont {Senthil}},\ }\href {https://doi.org/10.1103/kcm5-hx56} {\bibfield  {journal} {\bibinfo  {journal} {Phys. Rev. X}\ }\textbf {\bibinfo {volume} {15}},\ \bibinfo {pages} {031069} (\bibinfo {year} {2025}{\natexlab{a}})},\ \Eprint {https://arxiv.org/abs/2409.20567} {arXiv:2409.20567 [cond-mat.str-el]} \BibitemShut {NoStop}%
\bibitem [{\citenamefont {Schleith}\ \emph {et~al.}(2025)\citenamefont {Schleith}, \citenamefont {Soejima},\ and\ \citenamefont {Khalaf}}]{SchleithSoejimaKhalaf}%
  \BibitemOpen
  \bibfield  {author} {\bibinfo {author} {\bibfnamefont {M.-L.}\ \bibnamefont {Schleith}}, \bibinfo {author} {\bibfnamefont {T.}~\bibnamefont {Soejima}},\ and\ \bibinfo {author} {\bibfnamefont {E.}~\bibnamefont {Khalaf}},\ }\href {https://doi.org/10.48550/arXiv.2506.11211} {\bibinfo {title} {Anyon dispersion from non-uniform magnetic field on the sphere}} (\bibinfo {year} {2025}),\ \Eprint {https://arxiv.org/abs/2506.11211} {arXiv:2506.11211} \BibitemShut {NoStop}%
\bibitem [{\citenamefont {Iyer}\ \emph {et~al.}(2026)\citenamefont {Iyer}, \citenamefont {Feuerpfeil}, \citenamefont {Cr{\'e}pel}, \citenamefont {Regnault},\ and\ \citenamefont {Mora}}]{IyerAnyonBloch}%
  \BibitemOpen
  \bibfield  {author} {\bibinfo {author} {\bibfnamefont {K.}~\bibnamefont {Iyer}}, \bibinfo {author} {\bibfnamefont {A.}~\bibnamefont {Feuerpfeil}}, \bibinfo {author} {\bibfnamefont {V.}~\bibnamefont {Cr{\'e}pel}}, \bibinfo {author} {\bibfnamefont {N.}~\bibnamefont {Regnault}},\ and\ \bibinfo {author} {\bibfnamefont {C.}~\bibnamefont {Mora}},\ }\href {https://doi.org/10.48550/arXiv.2604.24859} {\bibinfo {title} {Dispersion of anyon {Bloch} bands}} (\bibinfo {year} {2026}),\ \Eprint {https://arxiv.org/abs/2604.24859} {arXiv:2604.24859} \BibitemShut {NoStop}%
\bibitem [{\citenamefont {Yan}\ \emph {et~al.}(2025)\citenamefont {Yan}, \citenamefont {Li}, \citenamefont {Soejima},\ and\ \citenamefont {Khalaf}}]{YanLiSoejimaKhalaf}%
  \BibitemOpen
  \bibfield  {author} {\bibinfo {author} {\bibfnamefont {Z.}~\bibnamefont {Yan}}, \bibinfo {author} {\bibfnamefont {Q.}~\bibnamefont {Li}}, \bibinfo {author} {\bibfnamefont {T.}~\bibnamefont {Soejima}},\ and\ \bibinfo {author} {\bibfnamefont {E.}~\bibnamefont {Khalaf}},\ }\href {https://doi.org/10.48550/arXiv.2512.15863} {\bibinfo {title} {Anyon dispersion in {Aharonov-Casher} bands and implications for twisted {MoTe$_2$}}} (\bibinfo {year} {2025}),\ \Eprint {https://arxiv.org/abs/2512.15863} {arXiv:2512.15863} \BibitemShut {NoStop}%
\bibitem [{\citenamefont {Lee}\ \emph {et~al.}(2018)\citenamefont {Lee}, \citenamefont {Wang}, \citenamefont {Zaletel}, \citenamefont {Vishwanath},\ and\ \citenamefont {He}}]{LeeWangZaletelVishwanathHe2018}%
  \BibitemOpen
  \bibfield  {author} {\bibinfo {author} {\bibfnamefont {J.~Y.}\ \bibnamefont {Lee}}, \bibinfo {author} {\bibfnamefont {C.}~\bibnamefont {Wang}}, \bibinfo {author} {\bibfnamefont {M.~P.}\ \bibnamefont {Zaletel}}, \bibinfo {author} {\bibfnamefont {A.}~\bibnamefont {Vishwanath}},\ and\ \bibinfo {author} {\bibfnamefont {Y.-C.}\ \bibnamefont {He}},\ }\href {https://doi.org/10.1103/PhysRevX.8.031015} {\bibfield  {journal} {\bibinfo  {journal} {Phys. Rev. X}\ }\textbf {\bibinfo {volume} {8}},\ \bibinfo {pages} {031015} (\bibinfo {year} {2018})},\ \Eprint {https://arxiv.org/abs/1802.09538} {arXiv:1802.09538} \BibitemShut {NoStop}%
\bibitem [{\citenamefont {Wang}\ \emph {et~al.}(2025)\citenamefont {Wang}, \citenamefont {Song}, \citenamefont {Zaletel},\ and\ \citenamefont {Senthil}}]{WangSongZaletelSenthil2025}%
  \BibitemOpen
  \bibfield  {author} {\bibinfo {author} {\bibfnamefont {T.}~\bibnamefont {Wang}}, \bibinfo {author} {\bibfnamefont {X.-Y.}\ \bibnamefont {Song}}, \bibinfo {author} {\bibfnamefont {M.~P.}\ \bibnamefont {Zaletel}},\ and\ \bibinfo {author} {\bibfnamefont {T.}~\bibnamefont {Senthil}},\ }\href {https://doi.org/10.48550/arXiv.2507.07611} {\bibinfo {title} {Emergent {{QED}$_3$} at the bosonic {Laughlin} state to superfluid transition}} (\bibinfo {year} {2025}),\ \Eprint {https://arxiv.org/abs/2507.07611} {arXiv:2507.07611} \BibitemShut {NoStop}%
\bibitem [{\citenamefont {Lu}\ \emph {et~al.}(2025{\natexlab{b}})\citenamefont {Lu}, \citenamefont {Wu}, \citenamefont {Chen},\ and\ \citenamefont {Meng}}]{LuWuChenMeng2025}%
  \BibitemOpen
  \bibfield  {author} {\bibinfo {author} {\bibfnamefont {H.}~\bibnamefont {Lu}}, \bibinfo {author} {\bibfnamefont {H.-Q.}\ \bibnamefont {Wu}}, \bibinfo {author} {\bibfnamefont {B.-B.}\ \bibnamefont {Chen}},\ and\ \bibinfo {author} {\bibfnamefont {Z.~Y.}\ \bibnamefont {Meng}},\ }\href {https://doi.org/10.1103/PhysRevLett.134.076601} {\bibfield  {journal} {\bibinfo  {journal} {Phys. Rev. Lett.}\ }\textbf {\bibinfo {volume} {134}},\ \bibinfo {pages} {076601} (\bibinfo {year} {2025}{\natexlab{b}})},\ \Eprint {https://arxiv.org/abs/2405.18269} {arXiv:2405.18269} \BibitemShut {NoStop}%
\bibitem [{\citenamefont {Han}\ \emph {et~al.}(2025)\citenamefont {Han}, \citenamefont {Lu}, \citenamefont {Hadjri}, \citenamefont {Shi}, \citenamefont {Wu}, \citenamefont {Xu}, \citenamefont {Yao}, \citenamefont {Cotten}, \citenamefont {Sharifi~Sedeh}, \citenamefont {Weldeyesus}, \citenamefont {Yang}, \citenamefont {Seo}, \citenamefont {Ye}, \citenamefont {Zhou}, \citenamefont {Liu}, \citenamefont {Shi}, \citenamefont {Hua}, \citenamefont {Watanabe}, \citenamefont {Taniguchi}, \citenamefont {Xiong}, \citenamefont {Zumb{\"u}hl}, \citenamefont {Fu},\ and\ \citenamefont {Ju}}]{HanRhomboSC}%
  \BibitemOpen
  \bibfield  {author} {\bibinfo {author} {\bibfnamefont {T.}~\bibnamefont {Han}}, \bibinfo {author} {\bibfnamefont {Z.}~\bibnamefont {Lu}}, \bibinfo {author} {\bibfnamefont {Z.}~\bibnamefont {Hadjri}}, \bibinfo {author} {\bibfnamefont {L.}~\bibnamefont {Shi}}, \bibinfo {author} {\bibfnamefont {Z.}~\bibnamefont {Wu}}, \bibinfo {author} {\bibfnamefont {W.}~\bibnamefont {Xu}}, \bibinfo {author} {\bibfnamefont {Y.}~\bibnamefont {Yao}}, \bibinfo {author} {\bibfnamefont {A.~A.}\ \bibnamefont {Cotten}}, \bibinfo {author} {\bibfnamefont {O.}~\bibnamefont {Sharifi~Sedeh}}, \bibinfo {author} {\bibfnamefont {H.}~\bibnamefont {Weldeyesus}}, \bibinfo {author} {\bibfnamefont {J.}~\bibnamefont {Yang}}, \bibinfo {author} {\bibfnamefont {J.}~\bibnamefont {Seo}}, \bibinfo {author} {\bibfnamefont {S.}~\bibnamefont {Ye}}, \bibinfo {author} {\bibfnamefont {M.}~\bibnamefont {Zhou}}, \bibinfo {author} {\bibfnamefont {H.}~\bibnamefont {Liu}}, \bibinfo {author} {\bibfnamefont {G.}~\bibnamefont {Shi}}, \bibinfo {author} {\bibfnamefont
  {Z.}~\bibnamefont {Hua}}, \bibinfo {author} {\bibfnamefont {K.}~\bibnamefont {Watanabe}}, \bibinfo {author} {\bibfnamefont {T.}~\bibnamefont {Taniguchi}}, \bibinfo {author} {\bibfnamefont {P.}~\bibnamefont {Xiong}}, \bibinfo {author} {\bibfnamefont {D.~M.}\ \bibnamefont {Zumb{\"u}hl}}, \bibinfo {author} {\bibfnamefont {L.}~\bibnamefont {Fu}},\ and\ \bibinfo {author} {\bibfnamefont {L.}~\bibnamefont {Ju}},\ }\href {https://doi.org/10.1038/s41586-025-09169-7} {\bibfield  {journal} {\bibinfo  {journal} {Nature}\ }\textbf {\bibinfo {volume} {643}},\ \bibinfo {pages} {654} (\bibinfo {year} {2025})},\ \Eprint {https://arxiv.org/abs/2408.15233} {arXiv:2408.15233} \BibitemShut {NoStop}%
\bibitem [{\citenamefont {Xu}\ \emph {et~al.}(2025{\natexlab{a}})\citenamefont {Xu}, \citenamefont {Sun}, \citenamefont {Li}, \citenamefont {Zheng}, \citenamefont {Xu}, \citenamefont {Gao}, \citenamefont {Jia}, \citenamefont {Su}, \citenamefont {Watanabe}, \citenamefont {Taniguchi}, \citenamefont {Tong}, \citenamefont {Lu}, \citenamefont {Jia}, \citenamefont {Shi}, \citenamefont {Jiang}, \citenamefont {Lin}, \citenamefont {Zhang}, \citenamefont {Zhang}, \citenamefont {Lei}, \citenamefont {Liu},\ and\ \citenamefont {Li}}]{XuMoTe2SC}%
  \BibitemOpen
  \bibfield  {author} {\bibinfo {author} {\bibfnamefont {F.}~\bibnamefont {Xu}}, \bibinfo {author} {\bibfnamefont {Z.}~\bibnamefont {Sun}}, \bibinfo {author} {\bibfnamefont {J.}~\bibnamefont {Li}}, \bibinfo {author} {\bibfnamefont {C.}~\bibnamefont {Zheng}}, \bibinfo {author} {\bibfnamefont {C.}~\bibnamefont {Xu}}, \bibinfo {author} {\bibfnamefont {J.}~\bibnamefont {Gao}}, \bibinfo {author} {\bibfnamefont {T.}~\bibnamefont {Jia}}, \bibinfo {author} {\bibfnamefont {Y.}~\bibnamefont {Su}}, \bibinfo {author} {\bibfnamefont {K.}~\bibnamefont {Watanabe}}, \bibinfo {author} {\bibfnamefont {T.}~\bibnamefont {Taniguchi}}, \bibinfo {author} {\bibfnamefont {B.}~\bibnamefont {Tong}}, \bibinfo {author} {\bibfnamefont {L.}~\bibnamefont {Lu}}, \bibinfo {author} {\bibfnamefont {J.}~\bibnamefont {Jia}}, \bibinfo {author} {\bibfnamefont {Z.}~\bibnamefont {Shi}}, \bibinfo {author} {\bibfnamefont {S.}~\bibnamefont {Jiang}}, \bibinfo {author} {\bibfnamefont {J.}~\bibnamefont {Lin}}, \bibinfo {author} {\bibfnamefont
  {Y.}~\bibnamefont {Zhang}}, \bibinfo {author} {\bibfnamefont {Y.}~\bibnamefont {Zhang}}, \bibinfo {author} {\bibfnamefont {S.}~\bibnamefont {Lei}}, \bibinfo {author} {\bibfnamefont {X.}~\bibnamefont {Liu}},\ and\ \bibinfo {author} {\bibfnamefont {T.}~\bibnamefont {Li}},\ }\href {https://doi.org/10.48550/arXiv.2504.06972} {\bibinfo {title} {Signatures of unconventional superconductivity near reentrant and fractional quantum anomalous {Hall} insulators}} (\bibinfo {year} {2025}{\natexlab{a}}),\ \Eprint {https://arxiv.org/abs/2504.06972} {arXiv:2504.06972 [cond-mat.mes-hall]} \BibitemShut {NoStop}%
\bibitem [{\citenamefont {Laughlin}(1988{\natexlab{a}})}]{LaughlinAnyonSC}%
  \BibitemOpen
  \bibfield  {author} {\bibinfo {author} {\bibfnamefont {R.~B.}\ \bibnamefont {Laughlin}},\ }\href {https://doi.org/10.1103/PhysRevLett.60.2677} {\bibfield  {journal} {\bibinfo  {journal} {Phys. Rev. Lett.}\ }\textbf {\bibinfo {volume} {60}},\ \bibinfo {pages} {2677} (\bibinfo {year} {1988}{\natexlab{a}})}\BibitemShut {NoStop}%
\bibitem [{\citenamefont {Laughlin}(1988{\natexlab{b}})}]{LaughlinScience1988}%
  \BibitemOpen
  \bibfield  {author} {\bibinfo {author} {\bibfnamefont {R.~B.}\ \bibnamefont {Laughlin}},\ }\href {https://doi.org/10.1126/science.242.4878.525} {\bibfield  {journal} {\bibinfo  {journal} {Science}\ }\textbf {\bibinfo {volume} {242}},\ \bibinfo {pages} {525} (\bibinfo {year} {1988}{\natexlab{b}})}\BibitemShut {NoStop}%
\bibitem [{\citenamefont {Chen}\ \emph {et~al.}(1989)\citenamefont {Chen}, \citenamefont {Wilczek}, \citenamefont {Witten},\ and\ \citenamefont {Halperin}}]{Chen1989_anyonSC}%
  \BibitemOpen
  \bibfield  {author} {\bibinfo {author} {\bibfnamefont {Y.-H.}\ \bibnamefont {Chen}}, \bibinfo {author} {\bibfnamefont {F.}~\bibnamefont {Wilczek}}, \bibinfo {author} {\bibfnamefont {E.}~\bibnamefont {Witten}},\ and\ \bibinfo {author} {\bibfnamefont {B.~I.}\ \bibnamefont {Halperin}},\ }\href {https://doi.org/10.1142/S0217979289000725} {\bibfield  {journal} {\bibinfo  {journal} {Int. J. Mod. Phys. B}\ }\textbf {\bibinfo {volume} {3}},\ \bibinfo {pages} {1001} (\bibinfo {year} {1989})}\BibitemShut {NoStop}%
\bibitem [{\citenamefont {Lee}\ and\ \citenamefont {Fisher}(1989)}]{LeeFisher1989}%
  \BibitemOpen
  \bibfield  {author} {\bibinfo {author} {\bibfnamefont {D.-H.}\ \bibnamefont {Lee}}\ and\ \bibinfo {author} {\bibfnamefont {M.~P.~A.}\ \bibnamefont {Fisher}},\ }\href {https://doi.org/10.1103/PhysRevLett.63.903} {\bibfield  {journal} {\bibinfo  {journal} {Phys. Rev. Lett.}\ }\textbf {\bibinfo {volume} {63}},\ \bibinfo {pages} {903} (\bibinfo {year} {1989})}\BibitemShut {NoStop}%
\bibitem [{\citenamefont {Halperin}(1992)}]{Halperin1992}%
  \BibitemOpen
  \bibfield  {author} {\bibinfo {author} {\bibfnamefont {B.~I.}\ \bibnamefont {Halperin}},\ }\href {https://doi.org/10.1103/PhysRevB.45.5504} {\bibfield  {journal} {\bibinfo  {journal} {Phys. Rev. B}\ }\textbf {\bibinfo {volume} {45}},\ \bibinfo {pages} {5504} (\bibinfo {year} {1992})}\BibitemShut {NoStop}%
\bibitem [{\citenamefont {Kim}\ \emph {et~al.}(2025)\citenamefont {Kim}, \citenamefont {Timmel}, \citenamefont {Ju},\ and\ \citenamefont {Wen}}]{Kim2024_anyonSC}%
  \BibitemOpen
  \bibfield  {author} {\bibinfo {author} {\bibfnamefont {M.}~\bibnamefont {Kim}}, \bibinfo {author} {\bibfnamefont {A.}~\bibnamefont {Timmel}}, \bibinfo {author} {\bibfnamefont {L.}~\bibnamefont {Ju}},\ and\ \bibinfo {author} {\bibfnamefont {X.-G.}\ \bibnamefont {Wen}},\ }\href {https://doi.org/10.1103/PhysRevB.111.014508} {\bibfield  {journal} {\bibinfo  {journal} {Phys. Rev. B}\ }\textbf {\bibinfo {volume} {111}},\ \bibinfo {pages} {014508} (\bibinfo {year} {2025})},\ \Eprint {https://arxiv.org/abs/2409.18067} {arXiv:2409.18067 [cond-mat.str-el]} \BibitemShut {NoStop}%
\bibitem [{\citenamefont {Shi}\ \emph {et~al.}(2025)\citenamefont {Shi}, \citenamefont {Zhang},\ and\ \citenamefont {Senthil}}]{shi2025doping}%
  \BibitemOpen
  \bibfield  {author} {\bibinfo {author} {\bibfnamefont {Z.~D.}\ \bibnamefont {Shi}}, \bibinfo {author} {\bibfnamefont {C.}~\bibnamefont {Zhang}},\ and\ \bibinfo {author} {\bibfnamefont {T.}~\bibnamefont {Senthil}},\ }\href {https://doi.org/10.21468/SciPostPhys.19.6.150} {\bibfield  {journal} {\bibinfo  {journal} {SciPost Phys.}\ }\textbf {\bibinfo {volume} {19}},\ \bibinfo {pages} {150} (\bibinfo {year} {2025})},\ \Eprint {https://arxiv.org/abs/2505.02893} {arXiv:2505.02893 [cond-mat.str-el]} \BibitemShut {NoStop}%
\bibitem [{\citenamefont {Shi}\ and\ \citenamefont {Senthil}(2025{\natexlab{b}})}]{shi2025anyon}%
  \BibitemOpen
  \bibfield  {author} {\bibinfo {author} {\bibfnamefont {Z.~D.}\ \bibnamefont {Shi}}\ and\ \bibinfo {author} {\bibfnamefont {T.}~\bibnamefont {Senthil}},\ }\href {https://doi.org/10.1073/pnas.2520608122} {\bibfield  {journal} {\bibinfo  {journal} {Proc. Natl. Acad. Sci. U.S.A.}\ }\textbf {\bibinfo {volume} {122}},\ \bibinfo {pages} {e2520608122} (\bibinfo {year} {2025}{\natexlab{b}})},\ \Eprint {https://arxiv.org/abs/2506.02128} {arXiv:2506.02128 [cond-mat.str-el]} \BibitemShut {NoStop}%
\bibitem [{\citenamefont {Nosov}\ \emph {et~al.}(2026)\citenamefont {Nosov}, \citenamefont {Han},\ and\ \citenamefont {Khalaf}}]{NosovHanKhalaf}%
  \BibitemOpen
  \bibfield  {author} {\bibinfo {author} {\bibfnamefont {P.~A.}\ \bibnamefont {Nosov}}, \bibinfo {author} {\bibfnamefont {Z.}~\bibnamefont {Han}},\ and\ \bibinfo {author} {\bibfnamefont {E.}~\bibnamefont {Khalaf}},\ }\href {https://doi.org/10.1103/6bgj-bfdn} {\bibfield  {journal} {\bibinfo  {journal} {Phys. Rev. Lett.}\ }\textbf {\bibinfo {volume} {136}},\ \bibinfo {pages} {106501} (\bibinfo {year} {2026})},\ \Eprint {https://arxiv.org/abs/2506.02108} {arXiv:2506.02108 [cond-mat.str-el]} \BibitemShut {NoStop}%
\bibitem [{\citenamefont {Pichler}\ \emph {et~al.}(2026)\citenamefont {Pichler}, \citenamefont {Kuhlenkamp}, \citenamefont {Knap},\ and\ \citenamefont {Vishwanath}}]{Pichler2025}%
  \BibitemOpen
  \bibfield  {author} {\bibinfo {author} {\bibfnamefont {F.}~\bibnamefont {Pichler}}, \bibinfo {author} {\bibfnamefont {C.}~\bibnamefont {Kuhlenkamp}}, \bibinfo {author} {\bibfnamefont {M.}~\bibnamefont {Knap}},\ and\ \bibinfo {author} {\bibfnamefont {A.}~\bibnamefont {Vishwanath}},\ }\href {https://doi.org/10.1016/j.newton.2025.100340} {\bibfield  {journal} {\bibinfo  {journal} {Newton}\ }\textbf {\bibinfo {volume} {2}},\ \bibinfo {pages} {100340} (\bibinfo {year} {2026})},\ \Eprint {https://arxiv.org/abs/2506.08000} {arXiv:2506.08000 [cond-mat.str-el]} \BibitemShut {NoStop}%
\bibitem [{\citenamefont {Wang}\ and\ \citenamefont {Zaletel}(2025)}]{WangZaletelSC}%
  \BibitemOpen
  \bibfield  {author} {\bibinfo {author} {\bibfnamefont {T.}~\bibnamefont {Wang}}\ and\ \bibinfo {author} {\bibfnamefont {M.~P.}\ \bibnamefont {Zaletel}},\ }\href {https://doi.org/10.48550/arXiv.2507.07921} {\bibinfo {title} {Chiral superconductivity near a fractional {Chern} insulator}} (\bibinfo {year} {2025}),\ \Eprint {https://arxiv.org/abs/2507.07921} {arXiv:2507.07921} \BibitemShut {NoStop}%
\bibitem [{\citenamefont {Fan}\ \emph {et~al.}(2026)\citenamefont {Fan}, \citenamefont {Vishwanath},\ and\ \citenamefont {Wang}}]{FanVishwanathWang2026}%
  \BibitemOpen
  \bibfield  {author} {\bibinfo {author} {\bibfnamefont {Z.-D.}\ \bibnamefont {Fan}}, \bibinfo {author} {\bibfnamefont {A.}~\bibnamefont {Vishwanath}},\ and\ \bibinfo {author} {\bibfnamefont {Z.}~\bibnamefont {Wang}},\ }\href {https://doi.org/10.48550/arXiv.2605.19036} {\bibinfo {title} {Hidden weak-pairing superconductivity of non-interacting anyons obeying {$1/3$} statistics}} (\bibinfo {year} {2026}),\ \Eprint {https://arxiv.org/abs/2605.19036} {arXiv:2605.19036 [cond-mat.str-el]} \BibitemShut {NoStop}%
\bibitem [{\citenamefont {Shi}\ and\ \citenamefont {Senthil}(2025{\natexlab{c}})}]{shi2025non}%
  \BibitemOpen
  \bibfield  {author} {\bibinfo {author} {\bibfnamefont {Z.~D.}\ \bibnamefont {Shi}}\ and\ \bibinfo {author} {\bibfnamefont {T.}~\bibnamefont {Senthil}},\ }\href {https://doi.org/10.48550/arXiv.2512.17996} {\bibinfo {title} {Non-{Abelian} topological superconductivity from melting {Abelian} fractional {Chern} insulators}} (\bibinfo {year} {2025}{\natexlab{c}}),\ \Eprint {https://arxiv.org/abs/2512.17996} {arXiv:2512.17996} \BibitemShut {NoStop}%
\bibitem [{\citenamefont {Crommie}\ \emph {et~al.}(1993)\citenamefont {Crommie}, \citenamefont {Lutz},\ and\ \citenamefont {Eigler}}]{Crommie1993}%
  \BibitemOpen
  \bibfield  {author} {\bibinfo {author} {\bibfnamefont {M.~F.}\ \bibnamefont {Crommie}}, \bibinfo {author} {\bibfnamefont {C.~P.}\ \bibnamefont {Lutz}},\ and\ \bibinfo {author} {\bibfnamefont {D.~M.}\ \bibnamefont {Eigler}},\ }\href {https://doi.org/10.1038/363524a0} {\bibfield  {journal} {\bibinfo  {journal} {Nature}\ }\textbf {\bibinfo {volume} {363}},\ \bibinfo {pages} {524} (\bibinfo {year} {1993})}\BibitemShut {NoStop}%
\bibitem [{\citenamefont {Hasegawa}\ and\ \citenamefont {Avouris}(1993)}]{Hasegawa1993}%
  \BibitemOpen
  \bibfield  {author} {\bibinfo {author} {\bibfnamefont {Y.}~\bibnamefont {Hasegawa}}\ and\ \bibinfo {author} {\bibfnamefont {P.}~\bibnamefont {Avouris}},\ }\href {https://doi.org/10.1103/PhysRevLett.71.1071} {\bibfield  {journal} {\bibinfo  {journal} {Phys. Rev. Lett.}\ }\textbf {\bibinfo {volume} {71}},\ \bibinfo {pages} {1071} (\bibinfo {year} {1993})}\BibitemShut {NoStop}%
\bibitem [{\citenamefont {Hoffman}\ \emph {et~al.}(2002)\citenamefont {Hoffman}, \citenamefont {McElroy}, \citenamefont {Lee}, \citenamefont {Lang}, \citenamefont {Eisaki}, \citenamefont {Uchida},\ and\ \citenamefont {Davis}}]{Hoffman2002}%
  \BibitemOpen
  \bibfield  {author} {\bibinfo {author} {\bibfnamefont {J.~E.}\ \bibnamefont {Hoffman}}, \bibinfo {author} {\bibfnamefont {K.}~\bibnamefont {McElroy}}, \bibinfo {author} {\bibfnamefont {D.-H.}\ \bibnamefont {Lee}}, \bibinfo {author} {\bibfnamefont {K.~M.}\ \bibnamefont {Lang}}, \bibinfo {author} {\bibfnamefont {H.}~\bibnamefont {Eisaki}}, \bibinfo {author} {\bibfnamefont {S.}~\bibnamefont {Uchida}},\ and\ \bibinfo {author} {\bibfnamefont {J.~C.}\ \bibnamefont {Davis}},\ }\href {https://doi.org/10.1126/science.1072640} {\bibfield  {journal} {\bibinfo  {journal} {Science}\ }\textbf {\bibinfo {volume} {297}},\ \bibinfo {pages} {1148} (\bibinfo {year} {2002})}\BibitemShut {NoStop}%
\bibitem [{\citenamefont {McElroy}\ \emph {et~al.}(2003)\citenamefont {McElroy}, \citenamefont {Simmonds}, \citenamefont {Hoffman}, \citenamefont {Lee}, \citenamefont {Orenstein}, \citenamefont {Eisaki}, \citenamefont {Uchida},\ and\ \citenamefont {Davis}}]{McElroy2003}%
  \BibitemOpen
  \bibfield  {author} {\bibinfo {author} {\bibfnamefont {K.}~\bibnamefont {McElroy}}, \bibinfo {author} {\bibfnamefont {R.~W.}\ \bibnamefont {Simmonds}}, \bibinfo {author} {\bibfnamefont {J.~E.}\ \bibnamefont {Hoffman}}, \bibinfo {author} {\bibfnamefont {D.-H.}\ \bibnamefont {Lee}}, \bibinfo {author} {\bibfnamefont {J.}~\bibnamefont {Orenstein}}, \bibinfo {author} {\bibfnamefont {H.}~\bibnamefont {Eisaki}}, \bibinfo {author} {\bibfnamefont {S.}~\bibnamefont {Uchida}},\ and\ \bibinfo {author} {\bibfnamefont {J.~C.}\ \bibnamefont {Davis}},\ }\href {https://doi.org/10.1038/nature01496} {\bibfield  {journal} {\bibinfo  {journal} {Nature}\ }\textbf {\bibinfo {volume} {422}},\ \bibinfo {pages} {592} (\bibinfo {year} {2003})}\BibitemShut {NoStop}%
\bibitem [{\citenamefont {Wang}\ and\ \citenamefont {Lee}(2003)}]{WangLee2003}%
  \BibitemOpen
  \bibfield  {author} {\bibinfo {author} {\bibfnamefont {Q.-H.}\ \bibnamefont {Wang}}\ and\ \bibinfo {author} {\bibfnamefont {D.-H.}\ \bibnamefont {Lee}},\ }\href {https://doi.org/10.1103/PhysRevB.67.020511} {\bibfield  {journal} {\bibinfo  {journal} {Phys. Rev. B}\ }\textbf {\bibinfo {volume} {67}},\ \bibinfo {pages} {020511} (\bibinfo {year} {2003})}\BibitemShut {NoStop}%
\bibitem [{\citenamefont {Capriotti}\ \emph {et~al.}(2003)\citenamefont {Capriotti}, \citenamefont {Scalapino},\ and\ \citenamefont {Sedgewick}}]{Capriotti2003}%
  \BibitemOpen
  \bibfield  {author} {\bibinfo {author} {\bibfnamefont {L.}~\bibnamefont {Capriotti}}, \bibinfo {author} {\bibfnamefont {D.~J.}\ \bibnamefont {Scalapino}},\ and\ \bibinfo {author} {\bibfnamefont {R.~D.}\ \bibnamefont {Sedgewick}},\ }\href {https://doi.org/10.1103/PhysRevB.68.014508} {\bibfield  {journal} {\bibinfo  {journal} {Phys. Rev. B}\ }\textbf {\bibinfo {volume} {68}},\ \bibinfo {pages} {014508} (\bibinfo {year} {2003})}\BibitemShut {NoStop}%
\bibitem [{\citenamefont {Pereg-Barnea}\ and\ \citenamefont {Franz}(2003)}]{PeregBarneaFranz2003}%
  \BibitemOpen
  \bibfield  {author} {\bibinfo {author} {\bibfnamefont {T.}~\bibnamefont {Pereg-Barnea}}\ and\ \bibinfo {author} {\bibfnamefont {M.}~\bibnamefont {Franz}},\ }\href {https://doi.org/10.1103/PhysRevB.68.180506} {\bibfield  {journal} {\bibinfo  {journal} {Phys. Rev. B}\ }\textbf {\bibinfo {volume} {68}},\ \bibinfo {pages} {180506} (\bibinfo {year} {2003})}\BibitemShut {NoStop}%
\bibitem [{\citenamefont {Hanaguri}\ \emph {et~al.}(2010)\citenamefont {Hanaguri}, \citenamefont {Niitaka}, \citenamefont {Kuroki},\ and\ \citenamefont {Takagi}}]{Hanaguri2010}%
  \BibitemOpen
  \bibfield  {author} {\bibinfo {author} {\bibfnamefont {T.}~\bibnamefont {Hanaguri}}, \bibinfo {author} {\bibfnamefont {S.}~\bibnamefont {Niitaka}}, \bibinfo {author} {\bibfnamefont {K.}~\bibnamefont {Kuroki}},\ and\ \bibinfo {author} {\bibfnamefont {H.}~\bibnamefont {Takagi}},\ }\href {https://doi.org/10.1126/science.1187399} {\bibfield  {journal} {\bibinfo  {journal} {Science}\ }\textbf {\bibinfo {volume} {328}},\ \bibinfo {pages} {474} (\bibinfo {year} {2010})}\BibitemShut {NoStop}%
\bibitem [{\citenamefont {Roushan}\ \emph {et~al.}(2009)\citenamefont {Roushan}, \citenamefont {Seo}, \citenamefont {Parker}, \citenamefont {Hor}, \citenamefont {Hsieh}, \citenamefont {Qian}, \citenamefont {Richardella}, \citenamefont {Hasan}, \citenamefont {Cava},\ and\ \citenamefont {Yazdani}}]{Roushan2009}%
  \BibitemOpen
  \bibfield  {author} {\bibinfo {author} {\bibfnamefont {P.}~\bibnamefont {Roushan}}, \bibinfo {author} {\bibfnamefont {J.}~\bibnamefont {Seo}}, \bibinfo {author} {\bibfnamefont {C.~V.}\ \bibnamefont {Parker}}, \bibinfo {author} {\bibfnamefont {Y.~S.}\ \bibnamefont {Hor}}, \bibinfo {author} {\bibfnamefont {D.}~\bibnamefont {Hsieh}}, \bibinfo {author} {\bibfnamefont {D.}~\bibnamefont {Qian}}, \bibinfo {author} {\bibfnamefont {A.}~\bibnamefont {Richardella}}, \bibinfo {author} {\bibfnamefont {M.~Z.}\ \bibnamefont {Hasan}}, \bibinfo {author} {\bibfnamefont {R.~J.}\ \bibnamefont {Cava}},\ and\ \bibinfo {author} {\bibfnamefont {A.}~\bibnamefont {Yazdani}},\ }\href {https://doi.org/10.1038/nature08308} {\bibfield  {journal} {\bibinfo  {journal} {Nature}\ }\textbf {\bibinfo {volume} {460}},\ \bibinfo {pages} {1106} (\bibinfo {year} {2009})}\BibitemShut {NoStop}%
\bibitem [{\citenamefont {Mallet}\ \emph {et~al.}(2012)\citenamefont {Mallet}, \citenamefont {Brihuega}, \citenamefont {Bose}, \citenamefont {Ugeda}, \citenamefont {G{\'o}mez-Rodr{\'i}guez}, \citenamefont {Kern},\ and\ \citenamefont {Veuillen}}]{Mallet2012}%
  \BibitemOpen
  \bibfield  {author} {\bibinfo {author} {\bibfnamefont {P.}~\bibnamefont {Mallet}}, \bibinfo {author} {\bibfnamefont {I.}~\bibnamefont {Brihuega}}, \bibinfo {author} {\bibfnamefont {S.}~\bibnamefont {Bose}}, \bibinfo {author} {\bibfnamefont {M.~M.}\ \bibnamefont {Ugeda}}, \bibinfo {author} {\bibfnamefont {J.~M.}\ \bibnamefont {G{\'o}mez-Rodr{\'i}guez}}, \bibinfo {author} {\bibfnamefont {K.}~\bibnamefont {Kern}},\ and\ \bibinfo {author} {\bibfnamefont {J.~Y.}\ \bibnamefont {Veuillen}},\ }\href {https://doi.org/10.1103/PhysRevB.86.045444} {\bibfield  {journal} {\bibinfo  {journal} {Phys. Rev. B}\ }\textbf {\bibinfo {volume} {86}},\ \bibinfo {pages} {045444} (\bibinfo {year} {2012})}\BibitemShut {NoStop}%
\bibitem [{\citenamefont {Simon}\ \emph {et~al.}(2011)\citenamefont {Simon}, \citenamefont {Bena}, \citenamefont {Vonau}, \citenamefont {Cranney},\ and\ \citenamefont {Aubel}}]{SimonReview}%
  \BibitemOpen
  \bibfield  {author} {\bibinfo {author} {\bibfnamefont {L.}~\bibnamefont {Simon}}, \bibinfo {author} {\bibfnamefont {C.}~\bibnamefont {Bena}}, \bibinfo {author} {\bibfnamefont {F.}~\bibnamefont {Vonau}}, \bibinfo {author} {\bibfnamefont {M.}~\bibnamefont {Cranney}},\ and\ \bibinfo {author} {\bibfnamefont {D.}~\bibnamefont {Aubel}},\ }\href {https://doi.org/10.1088/0022-3727/44/46/464010} {\bibfield  {journal} {\bibinfo  {journal} {J. Phys. D: Appl. Phys.}\ }\textbf {\bibinfo {volume} {44}},\ \bibinfo {pages} {464010} (\bibinfo {year} {2011})}\BibitemShut {NoStop}%
\bibitem [{\citenamefont {Bena}(2016)}]{BenaReview}%
  \BibitemOpen
  \bibfield  {author} {\bibinfo {author} {\bibfnamefont {C.}~\bibnamefont {Bena}},\ }\href {https://doi.org/10.1016/j.crhy.2015.11.006} {\bibfield  {journal} {\bibinfo  {journal} {C. R. Phys.}\ }\textbf {\bibinfo {volume} {17}},\ \bibinfo {pages} {302} (\bibinfo {year} {2016})}\BibitemShut {NoStop}%
\bibitem [{\citenamefont {Avraham}\ \emph {et~al.}(2018)\citenamefont {Avraham}, \citenamefont {Reiner}, \citenamefont {Kumar-Nayak}, \citenamefont {Morali}, \citenamefont {Batabyal}, \citenamefont {Yan},\ and\ \citenamefont {Beidenkopf}}]{AvrahamReview}%
  \BibitemOpen
  \bibfield  {author} {\bibinfo {author} {\bibfnamefont {N.}~\bibnamefont {Avraham}}, \bibinfo {author} {\bibfnamefont {J.}~\bibnamefont {Reiner}}, \bibinfo {author} {\bibfnamefont {A.}~\bibnamefont {Kumar-Nayak}}, \bibinfo {author} {\bibfnamefont {N.}~\bibnamefont {Morali}}, \bibinfo {author} {\bibfnamefont {R.}~\bibnamefont {Batabyal}}, \bibinfo {author} {\bibfnamefont {B.}~\bibnamefont {Yan}},\ and\ \bibinfo {author} {\bibfnamefont {H.}~\bibnamefont {Beidenkopf}},\ }\href {https://doi.org/10.1002/adma.201707628} {\bibfield  {journal} {\bibinfo  {journal} {Adv. Mater.}\ }\textbf {\bibinfo {volume} {30}},\ \bibinfo {pages} {1707628} (\bibinfo {year} {2018})}\BibitemShut {NoStop}%
\bibitem [{\citenamefont {Morampudi}\ \emph {et~al.}(2017)\citenamefont {Morampudi}, \citenamefont {Turner}, \citenamefont {Pollmann},\ and\ \citenamefont {Wilczek}}]{Morampudi2017Threshold}%
  \BibitemOpen
  \bibfield  {author} {\bibinfo {author} {\bibfnamefont {S.~C.}\ \bibnamefont {Morampudi}}, \bibinfo {author} {\bibfnamefont {A.~M.}\ \bibnamefont {Turner}}, \bibinfo {author} {\bibfnamefont {F.}~\bibnamefont {Pollmann}},\ and\ \bibinfo {author} {\bibfnamefont {F.}~\bibnamefont {Wilczek}},\ }\href {https://doi.org/10.1103/PhysRevLett.118.227201} {\bibfield  {journal} {\bibinfo  {journal} {Phys. Rev. Lett.}\ }\textbf {\bibinfo {volume} {118}},\ \bibinfo {pages} {227201} (\bibinfo {year} {2017})}\BibitemShut {NoStop}%
\bibitem [{\citenamefont {Inbar}\ \emph {et~al.}(2023)\citenamefont {Inbar}, \citenamefont {Birkbeck}, \citenamefont {Xiao}, \citenamefont {Taniguchi}, \citenamefont {Watanabe}, \citenamefont {Yan}, \citenamefont {Oreg}, \citenamefont {Stern}, \citenamefont {Berg},\ and\ \citenamefont {Ilani}}]{InbarQTM}%
  \BibitemOpen
  \bibfield  {author} {\bibinfo {author} {\bibfnamefont {A.}~\bibnamefont {Inbar}}, \bibinfo {author} {\bibfnamefont {J.}~\bibnamefont {Birkbeck}}, \bibinfo {author} {\bibfnamefont {J.}~\bibnamefont {Xiao}}, \bibinfo {author} {\bibfnamefont {T.}~\bibnamefont {Taniguchi}}, \bibinfo {author} {\bibfnamefont {K.}~\bibnamefont {Watanabe}}, \bibinfo {author} {\bibfnamefont {B.}~\bibnamefont {Yan}}, \bibinfo {author} {\bibfnamefont {Y.}~\bibnamefont {Oreg}}, \bibinfo {author} {\bibfnamefont {A.}~\bibnamefont {Stern}}, \bibinfo {author} {\bibfnamefont {E.}~\bibnamefont {Berg}},\ and\ \bibinfo {author} {\bibfnamefont {S.}~\bibnamefont {Ilani}},\ }\href {https://doi.org/10.1038/s41586-022-05685-y} {\bibfield  {journal} {\bibinfo  {journal} {Nature}\ }\textbf {\bibinfo {volume} {614}},\ \bibinfo {pages} {682} (\bibinfo {year} {2023})}\BibitemShut {NoStop}%
\bibitem [{\citenamefont {Xiao}\ \emph {et~al.}(2026)\citenamefont {Xiao}, \citenamefont {Inbar}, \citenamefont {Birkbeck}, \citenamefont {Gershon}, \citenamefont {Zamir}, \citenamefont {Vituri}, \citenamefont {Taniguchi}, \citenamefont {Watanabe}, \citenamefont {Berg},\ and\ \citenamefont {Ilani}}]{XiaoMATBGQTM}%
  \BibitemOpen
  \bibfield  {author} {\bibinfo {author} {\bibfnamefont {J.}~\bibnamefont {Xiao}}, \bibinfo {author} {\bibfnamefont {A.}~\bibnamefont {Inbar}}, \bibinfo {author} {\bibfnamefont {J.}~\bibnamefont {Birkbeck}}, \bibinfo {author} {\bibfnamefont {N.}~\bibnamefont {Gershon}}, \bibinfo {author} {\bibfnamefont {Y.}~\bibnamefont {Zamir}}, \bibinfo {author} {\bibfnamefont {Y.}~\bibnamefont {Vituri}}, \bibinfo {author} {\bibfnamefont {T.}~\bibnamefont {Taniguchi}}, \bibinfo {author} {\bibfnamefont {K.}~\bibnamefont {Watanabe}}, \bibinfo {author} {\bibfnamefont {E.}~\bibnamefont {Berg}},\ and\ \bibinfo {author} {\bibfnamefont {S.}~\bibnamefont {Ilani}},\ }\href {https://doi.org/10.1038/s41586-026-10378-x} {\bibfield  {journal} {\bibinfo  {journal} {Nature}\ }\textbf {\bibinfo {volume} {653}},\ \bibinfo {pages} {68} (\bibinfo {year} {2026})},\ \Eprint {https://arxiv.org/abs/2506.20738} {arXiv:2506.20738} \BibitemShut {NoStop}%
\bibitem [{\citenamefont {Wang}\ and\ \citenamefont {Zaletel}(2026)}]{WangZaletelMolecules}%
  \BibitemOpen
  \bibfield  {author} {\bibinfo {author} {\bibfnamefont {T.}~\bibnamefont {Wang}}\ and\ \bibinfo {author} {\bibfnamefont {M.~P.}\ \bibnamefont {Zaletel}},\ }\href {https://doi.org/10.48550/arXiv.2604.09798} {\bibinfo {title} {Anyon molecules in fractional quantum {Hall} states}} (\bibinfo {year} {2026}),\ \Eprint {https://arxiv.org/abs/2604.09798} {arXiv:2604.09798 [cond-mat.mes-hall]} \BibitemShut {NoStop}%
\bibitem [{\citenamefont {Xu}\ \emph {et~al.}(2025{\natexlab{b}})\citenamefont {Xu}, \citenamefont {Ji}, \citenamefont {Wang}, \citenamefont {Trung},\ and\ \citenamefont {Yang}}]{XuClusters}%
  \BibitemOpen
  \bibfield  {author} {\bibinfo {author} {\bibfnamefont {Q.}~\bibnamefont {Xu}}, \bibinfo {author} {\bibfnamefont {G.}~\bibnamefont {Ji}}, \bibinfo {author} {\bibfnamefont {Y.}~\bibnamefont {Wang}}, \bibinfo {author} {\bibfnamefont {H.~Q.}\ \bibnamefont {Trung}},\ and\ \bibinfo {author} {\bibfnamefont {B.}~\bibnamefont {Yang}},\ }\href {https://doi.org/10.1103/vgz6-z98r} {\bibfield  {journal} {\bibinfo  {journal} {Phys. Rev. B}\ }\textbf {\bibinfo {volume} {112}},\ \bibinfo {pages} {235112} (\bibinfo {year} {2025}{\natexlab{b}})}\BibitemShut {NoStop}%
\bibitem [{\citenamefont {Gattu}\ and\ \citenamefont {Jain}(2025)}]{GattuJainMolecular}%
  \BibitemOpen
  \bibfield  {author} {\bibinfo {author} {\bibfnamefont {M.}~\bibnamefont {Gattu}}\ and\ \bibinfo {author} {\bibfnamefont {J.~K.}\ \bibnamefont {Jain}},\ }\href {https://doi.org/10.1103/scl5-8pv6} {\bibfield  {journal} {\bibinfo  {journal} {Phys. Rev. Lett.}\ }\textbf {\bibinfo {volume} {135}},\ \bibinfo {pages} {236601} (\bibinfo {year} {2025})}\BibitemShut {NoStop}%
\bibitem [{\citenamefont {Li}\ \emph {et~al.}(2026)\citenamefont {Li}, \citenamefont {Nosov}, \citenamefont {Wang},\ and\ \citenamefont {Khalaf}}]{LiNosovKhalaf}%
  \BibitemOpen
  \bibfield  {author} {\bibinfo {author} {\bibfnamefont {Q.}~\bibnamefont {Li}}, \bibinfo {author} {\bibfnamefont {P.~A.}\ \bibnamefont {Nosov}}, \bibinfo {author} {\bibfnamefont {T.}~\bibnamefont {Wang}},\ and\ \bibinfo {author} {\bibfnamefont {E.}~\bibnamefont {Khalaf}},\ }\href {https://doi.org/10.48550/arXiv.2603.24701} {\bibinfo {title} {Bound states of anyons: A geometric quantization approach}} (\bibinfo {year} {2026}),\ \Eprint {https://arxiv.org/abs/2603.24701} {arXiv:2603.24701} \BibitemShut {NoStop}%
\bibitem [{\citenamefont {Pu}\ \emph {et~al.}(2024)\citenamefont {Pu}, \citenamefont {Balram}, \citenamefont {Hu}, \citenamefont {Tsui}, \citenamefont {He}, \citenamefont {Regnault}, \citenamefont {Zaletel}, \citenamefont {Yazdani},\ and\ \citenamefont {Papi{\'c}}}]{PuCompositeFermionSTM}%
  \BibitemOpen
  \bibfield  {author} {\bibinfo {author} {\bibfnamefont {S.}~\bibnamefont {Pu}}, \bibinfo {author} {\bibfnamefont {A.~C.}\ \bibnamefont {Balram}}, \bibinfo {author} {\bibfnamefont {Y.}~\bibnamefont {Hu}}, \bibinfo {author} {\bibfnamefont {Y.-C.}\ \bibnamefont {Tsui}}, \bibinfo {author} {\bibfnamefont {M.}~\bibnamefont {He}}, \bibinfo {author} {\bibfnamefont {N.}~\bibnamefont {Regnault}}, \bibinfo {author} {\bibfnamefont {M.~P.}\ \bibnamefont {Zaletel}}, \bibinfo {author} {\bibfnamefont {A.}~\bibnamefont {Yazdani}},\ and\ \bibinfo {author} {\bibfnamefont {Z.}~\bibnamefont {Papi{\'c}}},\ }\href {https://doi.org/10.1103/PhysRevB.110.L081107} {\bibfield  {journal} {\bibinfo  {journal} {Phys. Rev. B}\ }\textbf {\bibinfo {volume} {110}},\ \bibinfo {pages} {L081107} (\bibinfo {year} {2024})},\ \Eprint {https://arxiv.org/abs/2312.06779} {arXiv:2312.06779 [cond-mat.mes-hall]} \BibitemShut {NoStop}%
\end{thebibliography}%

\end{document}